\documentclass[journal,twoside,web] {ieeecolor}
\usepackage{jsen}
\usepackage{cite}
\usepackage{amsmath,amssymb,amsfonts}
\usepackage{algorithmic}
\usepackage{graphicx}
\usepackage{textcomp}
\usepackage{wrapfig}
\usepackage{subfig}
\usepackage{subcaption}

\def\BibTeX{{\rm B\kern-.05em{\sc i\kern-.025em b}\kern-.08em
    T\kern-.1667em\lower.7ex\hbox{E}\kern-.125emX}}
\definecolor{abstractbg}{rgb}{0.89804,0.94510,0.83137}
\begin{document}
\title{Interpretable and Efficient Beamforming-Based Deep Learning for Single Snapshot DOA Estimation}
\author{Ruxin Zheng, \IEEEmembership{Student Member, IEEE}, Shunqiao Sun, \IEEEmembership{Senior Member, IEEE}, Hongshan Liu, \IEEEmembership{Student Member, IEEE}, Honglei Chen, \IEEEmembership{Member, IEEE}, and Jian Li, \IEEEmembership{Fellow, IEEE}
\thanks{This work was
supported in part by U.S. National Science Foundation (NSF) under Grant CCF-2153386. }
\thanks{R. Zheng, S. Sun and H. Liu are with the Department of Electrical and Computer Engineering, The University of Alabama,  Tuscaloosa, AL 38457 USA (e-mails: rzheng9@crimson.ua.edu, shunqiao.sun@ua.edu, hliu75@crimson.ua.edu).}
\thanks{H. Chen is with Mathworks, Natick, MA 01760 USA (e-mail: hchen@mathworks.com).}
\thanks{J. Li is with the Department of Electrical and Computer Engineering, University of Florida, Gainesville, FL 32611 USA (e-mail: li@dsp.ufl.edu).}}

\IEEEtitleabstractindextext{%
\fcolorbox{abstractbg}{abstractbg}{%
\begin{minipage}{\textwidth}%
\begin{wrapfigure}[16]{r}{4.1in}%
\includegraphics[width =4in]{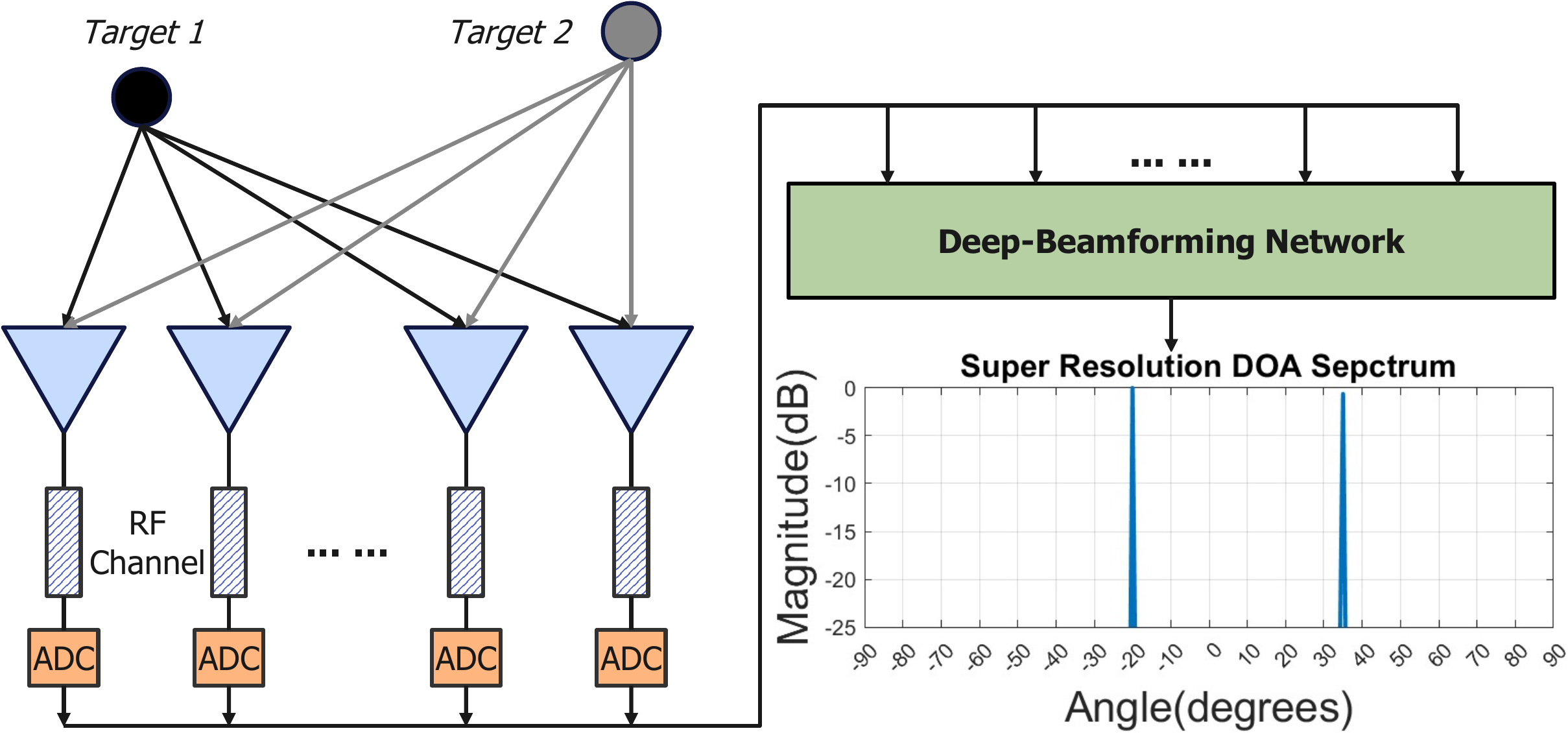}%
\end{wrapfigure}%
\begin{abstract}
We introduce an interpretable deep learning approach for direction of arrival (DOA) estimation with a single snapshot. Classical subspace-based methods like MUSIC and ESPRIT use spatial smoothing on uniform linear arrays for single snapshot DOA estimation but face drawbacks in reduced array aperture and inapplicability to sparse arrays. Single-snapshot methods such as compressive sensing and iterative adaptation approach (IAA) encounter challenges with high computational costs and slow convergence, hampering real-time use. Recent deep learning DOA methods offer promising accuracy and speed. However, the practical deployment of deep networks is hindered by their black-box nature. To address this, we propose a deep-MPDR network translating minimum power distortionless response (MPDR)-type beamformer into deep learning, enhancing generalization and efficiency. Comprehensive experiments conducted using both simulated and real-world datasets substantiate its dominance in terms of inference time and accuracy in comparison to conventional methods. Moreover, it excels in terms of efficiency, generalizability, and interpretability when contrasted with other deep learning DOA estimation networks.
\end{abstract}

\end{minipage}}}

\maketitle

\section{Introduction}
\label{sec:introduction}
\IEEEPARstart{D}{irection}-of-arrival (DOA) estimation, commonly referred to as direction finding, is a pivotal process in sensor array signal processing and various engineering fields such as radar, radio astronomy, sonar, navigation, remote sensing, wireless communications, biomedical engineering, and speech processing. Despite extensive study in the literature, where many algorithms have been proposed and their performances thoroughly analyzed, most of these efforts focus on the asymptotic scenario of a large number of snapshots. However,  in dynamically changing scenarios like those encountered in practical automotive radar applications, the available DOA estimation data is often constrained to only a limited number of radar sensor array snapshots or, in the most challenging situations, even a single snapshot\cite{Jian_07,Patole_SPM_2017,engels2017advances, SUN_SPM_Feature_Article_2020, sun20214d, Markel_book_2022}.

The exploration of DOA estimation methods spans a significant historical trajectory\cite{pesavento2023three}. The conventional (Bartlett) beamformer, dating back to World War II, utilizes Fourier-based spectral analysis on spatio-temporally sampled data, but it suffered from high side lobe levels and limitations due to the Rayleigh resolution. Subsequently, the minimum power distortionless response (MPDR) beamformer and the minimum variance distortionless response (MVDR) beamformer, often referred to as the Capon beamformer, were introduced \cite{capon1969high,van2002detection}. These techniques aim to enhance source estimation in scenarios with closely spaced sources. The MPDR minimizes its output power under the constraint that the target signal is distortionless in the output, while the MVDR prioritizes signal power in the specified direction and simultaneously suppresses interference and noise from other angles.

Beyond beamforming methods, parametric subspace-based approaches, including techniques like Multiple Signal Classification (MUSIC)\cite{schmidt1982signal} and the Estimation of Parameters by Rotational Invariant Techniques (ESPRIT)\cite{Kailath_ESPRIT_1989}, along with their respective variants \cite{Liao_Single_Snapshot_MUSIC_2016, Haardt_Unitary_ESPRIT_1995}, retrieve DOA from data second-order statistics \cite{pisarenko1973retrieval}. DOA estimation can also be achieved using the nonlinear least squares (NLS) method, often referred to as the deterministic maximum likelihood (DML) estimation. DML typically requires a multidimensional grid search across the parameter space to determine the global minimum. Evidently, the computational complexity of this exhaustive multidimensional search strategy escalates exponentially with the number of sources \cite{pesavento2023three}. 

To mitigate the computational burden associated with NLS optimization, convex approximation methods rooted in sparse regularization have been proposed. Techniques based on compressive sensing (CS) \cite{donoho2006compressed}, which exploit the sparse nature of targets in the angular domain, have demonstrated super-resolution performance \cite{Candes_Super_Resolution_2014} and work effectively with single snapshots. For CS-based DOA estimation algorithms, the dictionary must satisfy the restricted isometry property (RIP) condition \cite{candes2007sparsity}, which demands an optimized design of antenna arrays to maintain low peak sidelobe levels \cite{SUN_SPM_Feature_Article_2020}. Another notable DOA estimation algorithm compatible with single snapshots is the iterative adaptive approach (IAA) \cite{Yardibi_IAA_2010, Roberts_IAA_2010}, employing an iterative and nonparametric approach. IAA has shown robustness in DOA estimation in comparison to CS-based methods.

However, these methods are subject to well-known limitations. Subspace-based techniques and the nonlinear least squares (NLS) method require prior knowledge of the source number, which might be challenging to obtain. Covariance-based methods like Capon's beamformer, MUSIC, and ESPRIT rely on a sufficient number of data snapshots to accurately estimate the data covariance matrix and can be affected by source correlations that lead to a rank deficiency in the sample data covariance matrix. Although spatial smoothing can alleviate some of these challenges by generating a smaller averaged covariance matrix, It is important to note that this technique is applicable only to uniform linear arrays and is not suitable for sparse arrays. Additionally, spatial smoothing results in a reduction of the effective aperture size of the array. In addition to these considerations, It is worth noting that super-resolution methods often entail substantial computational expense, requiring procedures like singular value decomposition (SVD), eigenvalue decomposition, matrix inversions on covariance matrices, or angle scanning.

In recent times, data-driven deep learning (DL) approaches for DOA estimation have gained significant traction \cite{papageorgiou2021deep,fuchs2022machine,feintuch2023neural,gall2020spectrum,gall2020learning}. Generally, DL-based methods offer several noteworthy advantages over traditional approaches, including rapid inference times and improved super-resolution capabilities \cite{papageorgiou2021deep}. However, It is important to acknowledge that deep learning techniques are predominantly data-driven and often lack interpretability. On the other hand, model-based deep learning methods \cite{shlezinger2022model, shlezinger2023model} aim to bridge this gap by combining the strengths of traditional mathematical models with data-driven systems. These approaches harness domain knowledge and mathematical structures tailored to specific problems, providing a more principled and interpretable framework while benefiting from limited data.  Some model-based deep learning techniques proposed in earlier research \cite{10052106, Zheng_EUSIPCO_2023} introduce a new category of robust DOA estimation solutions that effectively integrate available domain expertise. Nonetheless, their interpretability and performance with unseen array structures and an unknown number of sources are still constrained by their deep-learning nature. Consequently, the quest for interpretable, generalizable, and high-performance deep architectures in the realm of signal processing remains a crucial and ongoing challenge. 

In this paper, we present an interpretable and efficient deep learning network called deep-MPDR, which maps MPDR beamformer principles to a deep learning framework. MPDR beamformer is inherently interpretable as it leverages domain knowledge to model physical processes. Our approach enhances interpretability compared to conventional deep neural networks by emulating MPDR beamformer characteristics. Through comprehensive experiments utilizing simulated and real-world datasets across diverse signal-to-noise ratio (SNR) scenarios, we illustrate the superiority of deep-MPDR over traditional algorithms in both inference time and DOA estimation accuracy. Furthermore, deep-MPDR surpasses data-driven DL methods in terms of parameter efficiency, and generalization capability. These findings underscore the considerable potential of deep-MPDR as a promising solution for DOA estimation challenges, offering enhanced performance and interpretability compared to existing techniques.

\section{System Model}
In this section, we present the formulation of the DOA estimation problem. Additionally, we introduce the MPDR beamformer, and in conjunction, present the IAA algorithm.

\subsection{Signal Model}
Consider a scenario with $K$ narrowband far-field source signals $s_k$ for $k = 1,\cdots, K$, impinging on a general linear omnidirectional antenna array comprised of $N$ elements from direction $\theta_k$ for $k=1,\cdots, K$. The temporal differences among the sensors can be accurately captured through simple phase shifts, resulting in the following data model:
\begin{equation}
    \begin{aligned}\label{sig}
    \mathbf{y}(t) &= \sum_{k=1}^{K} \textbf{a}(\theta_k)s_k(t) +\textbf{n}(t)\\
        & = \textbf{A}( \boldsymbol{\theta} ) \textbf{s}(t) +\textbf{n}(t), \quad t = 1,\cdots, T,
    \end{aligned}
\end{equation}
where $t$ indexes the snapshot, $T$ is the number of snapshots, $\bf{n}$ represents a complex $N \times 1$  white Gaussian noise vector, and ${\bf A}(\boldsymbol{\theta})= \left [{\bf a}(\theta_1), {\bf a}(\theta_2),\cdots, {\bf a}(\theta_K) \right]$ is the $N \times K$ array manifold matrix, where
\begin{align}
 {\textbf a}(\theta) = \left [1, e^{\frac{2\pi d_2}{\lambda}\sin{\theta}}, \cdots , e^{\frac{2\pi d_{N}}{\lambda}\sin{\theta}} \right ]^T.
\end{align}
Here, $d_n$ is the element spacing between the $n$-th element and the first element, and ${\textbf{s}(t)} = [s_1(t),s_2(t), \cdots, s_K(t)]^T$ is the source vector. In this paper, we are interested in estimating the parameter $\theta$, i.e., the target DOAs, using a single snapshot of the array response $\bf{y}$. Accordingly, with $T$ equating to $1$, the signal snapshot model can be rephrased as $ \textbf{y} = \textbf{A}(\boldsymbol{\theta})\textbf{s}+\mathbf{n}$.

\begin{figure*}
\centering
\includegraphics[width= 6.05 in]{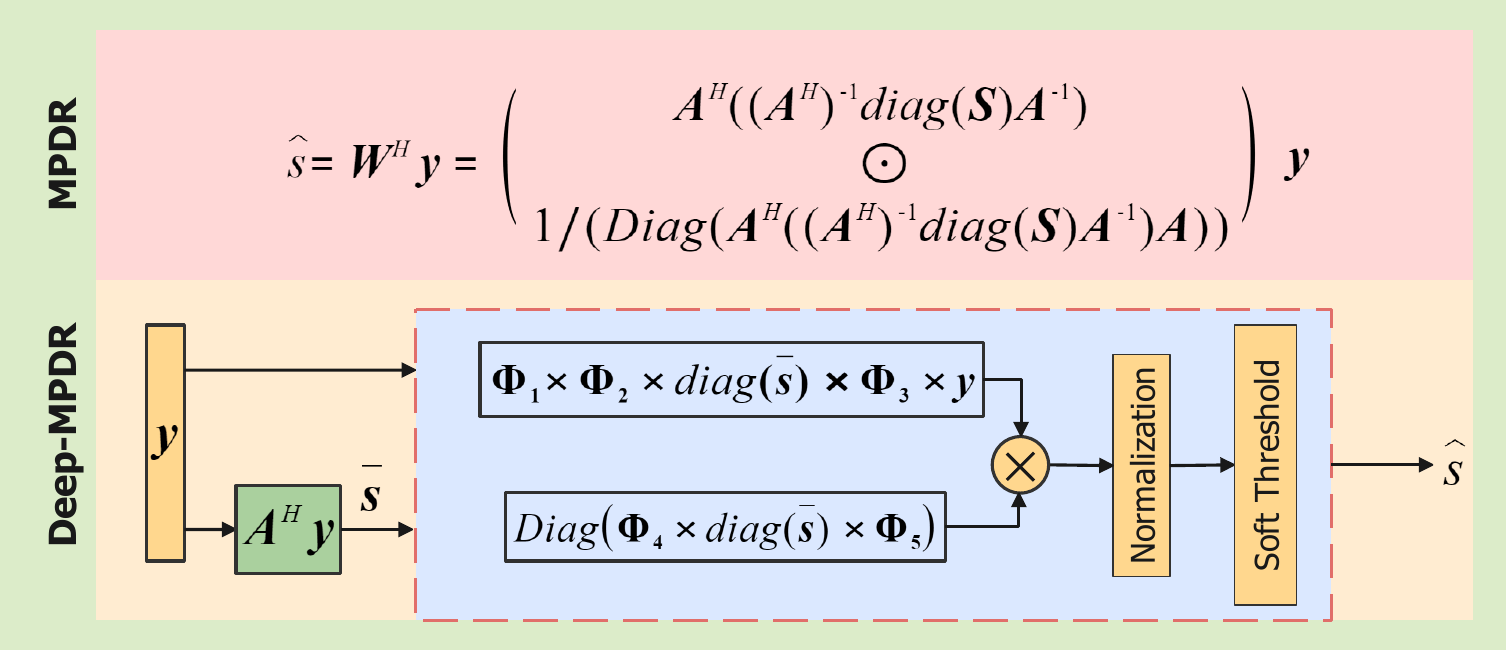}
\vspace{0mm}
\caption{The top section illustrates the MPDR-type beamformer, with ${\bf W}^H$ representing beamformer weights $\in \mathbb{C}^{L \times N}$ and $\bf{A}$ the array manifold matrix $\in \mathbb{C}^{N \times L}$. In the bottom section, we depict the deep-MPDR architecture. The symbol $\bf{\Phi}$ denotes a matrix of complex, learnable parameters within our custom deep learning layer, performing matrix multiplication with the input. Specifically, $\bf{\Phi}_1$ and $\bf{\Phi}_3$ $\in \mathbb{C}^{L \times N}$ correspond to the first and third layers, while $\bf{\Phi}_2$ $\in \mathbb{C}^{N \times L}$ represents the second layer. Additionally, $\bf{\Phi}_4$ and $\bf{\Phi}_5$ $\in \mathbb{C}^{L \times L}$ correspond to the fourth and fifth layers.}
\label{uaa_arch}

\end{figure*}

\subsection{MPDR \& IAA}

The MPDR beamformer is obtained through the minimization of power at the beamformer's output while ensuring that the signal from the intended direction remains undistorted. This can be formulated mathematically as follows
\begin{align}
{\textbf{w}_{\rm MPDR}} = \arg \min_{\textbf{w}} 
{ \textbf{w}^{H}\textbf{R} \textbf{w}}, \label{MPDR}
\end{align}
subjects to the constraint
\begin{align}
\textbf{w}^{H} \textbf{a}(\theta) = 1.
\end{align}
Here, $\textbf{w}^H$ denotes the Hermitian transpose of MPDR beamformer weights, and $\textbf{R}$ corresponds to the covariance matrix, which can be acquired using $\mathbf{R} = \mathbb{E}[\mathbf{yy}^H]$.
Incorporating the constraint into the objective function using a Lagrange multiplier and subsequently taking the complex gradient with respect to $\textbf{w}$, then setting the result to zero, leads to the solution for Equation (\ref{MPDR}) as:
\begin{align}
\textbf{w}_{\text{MPDR}}(\theta) = \frac{{\bf R}^{-1}{\bf a}(\theta)}{{\bf a}^H(\theta){\bf R}^{-1}{\bf a}(\theta)}.
\end{align}

The calculation of MPDR beamformer weights involves inverting the covariance matrix. However, accurately estimating the covariance matrix from a single snapshot presents challenges and results in performance degradation for the MPDR beamformer. 

IAA  is a data-dependent, nonparametric algorithm \cite{Yardibi_IAA_2010}. It discretizes the DOA space into an $L$ point grid, defining the array manifold as ${\bf A}(\boldsymbol{\theta}) = \left [{\bf a}(\theta_1), \cdots, {\bf a}(\theta_L) \right]$. The fictitious covariance matrix of $\bf{y}$ is represented as ${\bf R}_{f} = {\bf A}(\boldsymbol{\theta}){\bf P}{\bf A}^{H}(\boldsymbol{\theta})$, where $\bf{P}$ is an $L \times L$ diagonal matrix with the $l$-th diagonal element being $P_l = \frac{1}{T} \sum_{t=1}^{T}|{\hat s}_l(t)|^2$, and ${\hat s}_l$ is the source reflection coefficient corresponding to direction $\theta_l$.

IAA iteratively estimates the reflection coefficient $\hat{s}$ and updates the fictitious covariance matrix by minimizing the weighted least-square (WLS) cost function $\lVert {\bf y}-s_{l}{\bf a}(\theta_l) \rVert_{{\bf Q}^{-1}(\theta_l)}^{2}$, where $\lVert {\bf X} \rVert_{{\bf Q}^{-1}(\theta_l)}^{2} \overset{\Delta}{=} {\bf X}^{H}{\bf Q}^{-1}(\theta_l){\bf X}$ and ${\bf Q}(\theta_l) = {\bf R}_f-P_l{\bf a}(\theta_l){\bf a}^{H}(\theta_l)$. The solution to this optimization problem is 

\begin{equation}
\begin{aligned}\label{IAA}
{\hat s}_l &= \frac{{\bf a}^H(\theta_l){\bf R}_f^{-1}}{{\bf a}^H(\theta_l){\bf R}_f^{-1}{\bf a}(\theta_l)} {\bf y}. 
\end{aligned}
\end{equation}

From Equation (\ref{IAA}), each iteration of the IAA essentially involves performing MPDR type of beamforming. In IAA, new beamformer weights are iteratively calculated, coupled with the application of diagonal loading techniques \cite{li2003robust,vorobyov2003robust} to the fictitious covariance matrix before matrix inversion, which enhances the algorithm's robustness and stability.
Inspired by IAA, we can reformulate the calculation of MPDR-type beamformer weights for the single snapshot scenarios as follows
\begin{equation}
\begin{aligned} \label{newMPDR1}
\mathbf{w}^{H}(\theta) &= \frac{\mathbf{a}^H(\theta)\mathbf{R}_f^{-1}}{\mathbf{a}^H(\theta)\mathbf{R}_f^{-1}\mathbf{a}(\theta)} \\
&= \frac{\mathbf{a}^H(\theta)(\mathbf{A} \mathbf{P} \mathbf{A}^{H})^{-1}}{\mathbf{a}^H(\theta)(\mathbf{A} \mathbf{P} \mathbf{A}^{H})^{-1}\mathbf{a}(\theta)}.
\end{aligned}
\end{equation}
If $\mathbf{A}$ and $\mathbf{P}$ are invertible, we can write that,
\begin{equation}
\begin{aligned} \label{newMPDR2}
\mathbf{w}^{H}(\theta) &= \frac{\mathbf{a}^H(\theta)((\mathbf{A}^H)^{-1} \mathbf{P}^{-1} \mathbf{A}^{-1})}{\mathbf{a}^H(\theta)((\mathbf{A}^H)^{-1} \mathbf{P}^{-1} \mathbf{A}^{-1})\mathbf{a}(\theta)}.
\end{aligned}
\end{equation}
Furthermore, the beamformer weight for all $\theta$ can be written as
\begin{equation}
\begin{aligned} \label{newMPDR3}
\mathbf{W}^{H} = \mathbf{T}(\mathcal{S}) \odot \frac{1}{\mathbf{b}(\mathcal{S})},
\end{aligned}
\end{equation}
where
\begin{align} 
\mathbf{T}(\mathcal{S})  &= \mathbf{A}^H((\mathbf{A}^H)^{-1} \text{diag}(\mathcal{S}) \mathbf{A}^{-1}), \label{newMPDR4}\\
\mathbf{b}(\mathcal{S}) &= \text{Diag}(\mathbf{A}^H((\mathbf{A}^H)^{-1} \text{diag}(\mathcal{S}) \mathbf{A}^{-1}) \mathbf{A}).\label{newMPDR5}
\end{align}
Here, $ {\mathcal S } = \left[ 1/{|\bar s_1|}^2, 1/{|\bar s_2|}^2,  \cdots 1/{|\bar s_l|}^2 \right]$ and $\mathbf{\bar s} = \mathbf{A}^H \mathbf{y}$. The notation ${\rm diag}( \cdot )$ denotes the operation of creating a diagonal matrix using a vector, and ${\rm Diag}( \cdot )$ signifies extracting the diagonal elements of a matrix. $\bf{T}(\mathcal S)$ is an $L \times N$ complex-valued matrix, and $\bf{b}(\mathcal S)$ is an $L \times 1$ complex-valued vector. The symbol $\odot$ denotes row-wise matrix-vector multiplication. Specifically, the $i$-th row of $\mathbf{W}^{H}$ is computed by multiplying the $i$-th row of $\mathbf{T}(\mathcal S)$ by the $i$-th element of $1/\mathbf{b}(\mathcal{S})$. However, in most cases, $L \gg N$, which results in $\mathbf{A}$ being a fat matrix, and $\mathbf{A}^{-1}$ may not exist. Equations (\ref{newMPDR2}) through (\ref{newMPDR5}) are deliberately formulated in their current state to improve understanding of the mapping process that connects the MPDR-type beamformer, similar to a single iteration of IAA, with the architecture of deep learning networks, as explained in the following section. Henceforth,  we use the term `MPDR' to refer to our custom-defined MPDR-type beamformer for simplicity.

\section{Deep-MPDR Network For DOA Estimation}
In this section, we introduce the innovative deep-MPDR network, which integrates the conventional MPDR beamformer with deep networks. The deep-MPDR generates a pseudo-spectrum, containing estimated reflection power, which enables us to formulate the DOA estimation as a spectrum estimation problem, rather than a multi-label classification task. 

\subsection{The deep-MPDR Architecture}\label{arch}
We provide a comprehensive overview of the deep-MPDR architecture, highlighting its key components and innovative features. The fundamental concept driving deep-MPDR is the transformation of Equation (\ref{newMPDR3}) into a matrix multiplication involving multiple learnable parameter matrices with $\bf{\bar s}$. The procedure of mapping the MPDR beamformer to deep networks can be expressed as follows
\begin{equation}
\begin{aligned} \label{deepMPDR}
\textbf{W}^{H}_{\text{deepMPDR}}(\bf{\bar s}) = \bf{\hat T}(\bf{\bar s}) \odot  \bf{\hat b}(\bf{\bar s}),
\end{aligned}
\end{equation}
where, 
\begin{align}
\bf{\hat T}(\bf{\bar s}) = { {\mathbf{\Phi}_{1} \mathbf{\Phi}_{2} {\rm diag}(\bf{\bar s}) \mathbf{\Phi}_{3}}},
\end{align}
\begin{align}
\bf{\hat b}(\bf{\bar s}) = {{\rm Diag}( \mathbf{\Phi}_{4} {\rm diag}(\bf{\bar s}) \mathbf{\Phi}_{5})}.
\end{align}
Here, $\mathbf{\Phi}$ represents the learnable parameters matrix of our customized deep learning layer. This layer conducts matrix multiplication between $\mathbf{\Phi}$ and the input matrix.

The mapping between the deep-MPDR architecture and Equation (\ref{newMPDR3}) is illustrated in Fig. \ref{uaa_arch}. Here, $\bf{\hat T}(\bf{\bar s})$ and $\bf{\hat b}(\bf{\bar s})$ correspond to the numerator and denominator of the MPDR beamformer, respectively, with the reciprocal of $\mathbf{b}(\mathcal{S})$ being replaced by $\bf{\hat b}(\bf{\bar s})$. In other words, the $i$-th row of $\textbf{W}^{H}$ is obtained by the product of the $i$-th row of $\bf{\hat T}(\bf{\bar s})$ and the $i$-th element of $\bf{\hat b}(\bf{\bar s})$. Additionally, $\boldsymbol{\Phi}_1$ maps to the dictionary matrix, denoted as $\textbf{A}$. For simplicity, instead of taking $\mathcal{S}$ as input, we directly use $\bar {\bf s}$ as input and utilize $\boldsymbol{\Phi}_2$ and $\boldsymbol{\Phi}_3$ to perform operations akin to the inversion of the covariance matrix. Meanwhile, $\boldsymbol{\Phi}_4$ and $\boldsymbol{\Phi}_5$ are associated with diagonal-dominant matrices, with $\boldsymbol{\Phi}_4$ related to $\textbf{A}^H (\textbf{A}^H)^{-1}$ and $\boldsymbol{\Phi}_5$ linked to $\textbf{A}^{-1} \textbf{A}$.

Similar to the conventional MPDR beamformer, the spectrum estimation of the deep-MPDR, $\tilde{\bf s}$, is achieved as follows 
\begin{equation}
\begin{aligned} \label{MPDRy}
{\bf \tilde{s}} = {\mathbf W}_{\text{deepMPDR}}^H \mathbf{y}.
\end{aligned}
\end{equation}
The obtained $\tilde{ \bf s}$ is subsequently processed through a normalization layer followed by a soft threshold layer, resulting in the final estimation, $\bar{\bf s}$. The normalization layer ensures the input is scaled within the range of 0 to 1. On the other hand, the soft threshold layer incorporates two learnable parameters, $\alpha$ and $\beta$, defining the operation of soft thresholding as follows
\begin{equation}
\begin{aligned} \label{softThres}
\bf{\hat{s}} = \bf{\tilde{s}} * {\rm Sigmoid}\left (\alpha (\bf{\tilde{s}}-\beta ({\bf 1})) \right ),
\end{aligned}
\end{equation}
where $\bf 1$ is a vector containing all 1 entries, and $*$ stand for element-wise multiplication. Transforming the classical beamformer into a deep learning network represents a fusion of the strengths inherent in both paradigms. By harnessing the advantages of deep learning networks, such as exceptional performance in complex tasks, and integrating them with the interpretability offered by classical model-based approaches, we arrive at an efficient and accessible solution for DOA estimation. This unique synthesis yields superior performance and empowers us with a highly explainable and effective solution.

\subsection{Data Generation and Labeling}\label{data}
\subsubsection{Simulated Dataset}
We deploy a uniform linear array (ULA) consisting of $N=64$ elements with an inter-element spacing equivalent to half a wavelength. This configuration forms the basis for generating simulated beam vectors that depict $3$ targets, each identified by its DOA denoted as $\theta_k$, and maintaining a minimum separation of $\Delta \phi = 1^{\circ}$.
The radar's field of view (FOV), encompassing $\boldsymbol{\phi}_{\rm FOV} = [-90^{\circ},90^{\circ}]$, undergoes discretization in steps of $1^{\circ}$. The outcome is a grid ${\bf{g}} = {\left[ {{g_1}, \cdots ,{g_M}} \right]^T} \in {\mathbb R}^{M\times 1}$ where $M = 181$ potential DOA angles emerge. Each DOA source's reflection coefficients $s_k$ materialize as randomly generated complex numbers.

The label assigned to the beam vector is denoted as ${\bf{\hat g}} = {\left[ {{\hat g_1}, \cdots ,{\hat g_M}} \right]^T} \in {\mathbb R}^{M\times 1}$, and its form can be represented as:

\begin{equation}\label{bvgt}
  {\hat g_m} =
    \begin{cases}
      |s_k|, & \text{if } \theta_k = g_m\\
      0, & \text{else}
    \end{cases} 
\end{equation}

To fabricate these beam vectors, we amalgamate an array of diverse angles with randomly chosen reflection coefficients. The outcome is a set of $1,000,000$ beam vectors, each generated to maintain an SNR of $15$ dB, and all simulated targets are positioned on the angle grid. We opt for 3 targets when creating the simulated dataset due to the characteristics of automotive radars equipped with mmwave technology. These radars leverage a wide bandwidth, leading to high-range resolution. Consequently, only a limited number of targets fall within the same range-Doppler bin\cite{SUN_SPM_Feature_Article_2020}, making $3$ targets sufficient in such scenarios.

\begin{figure}
\centering
\includegraphics[width= \linewidth]{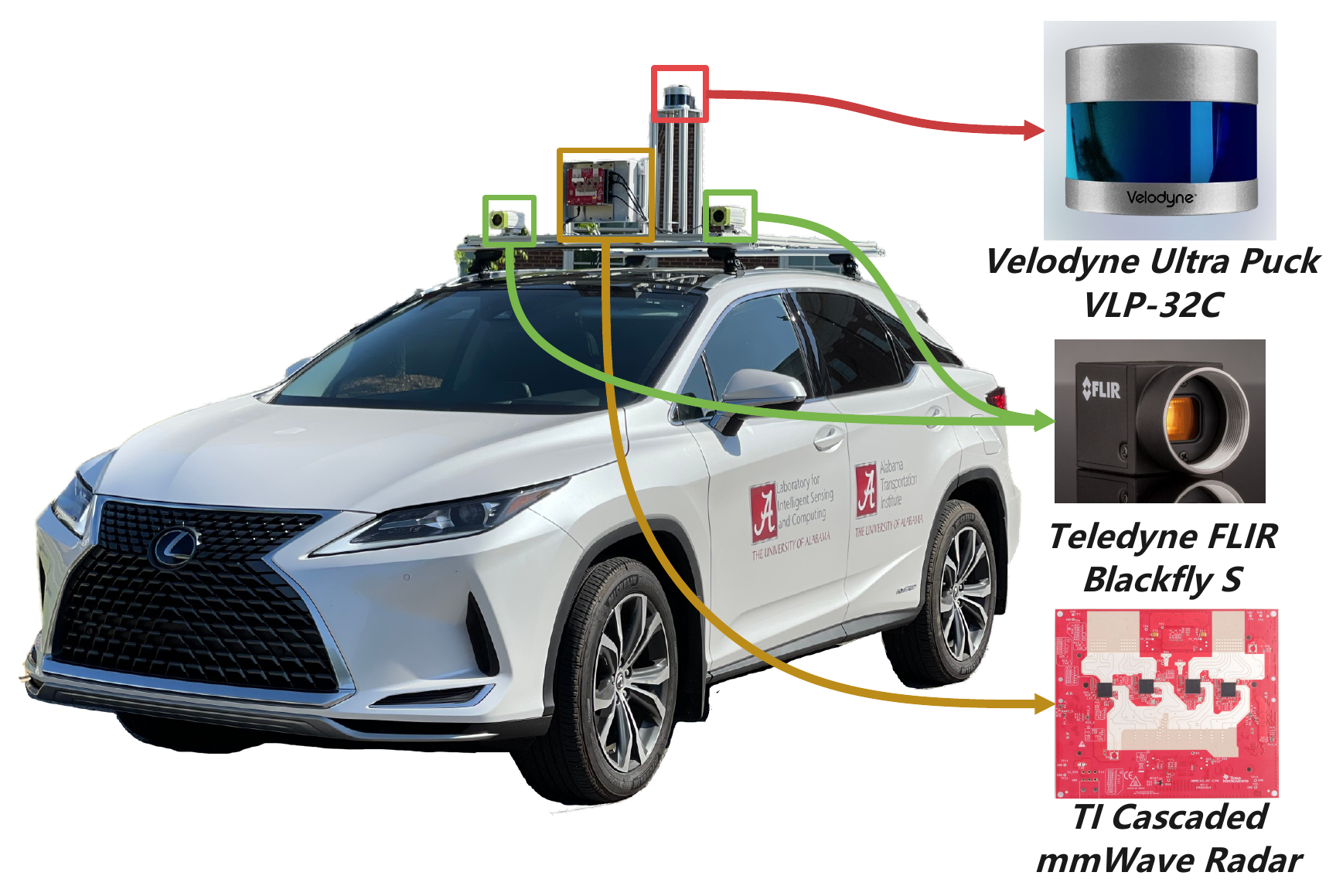}
\caption{  The data acquisition vehicle platform of Lexus RX450 Hybrid with high-resolution imaging radar, LiDAR, and stereo cameras, is used to carry out field experiments at The University of Alabama \cite{Ruxin_TAES_2023,zheng2023time}. 
} 
\label{plate}
\end{figure}

\begin{figure}
\centering
\includegraphics[width= \linewidth]{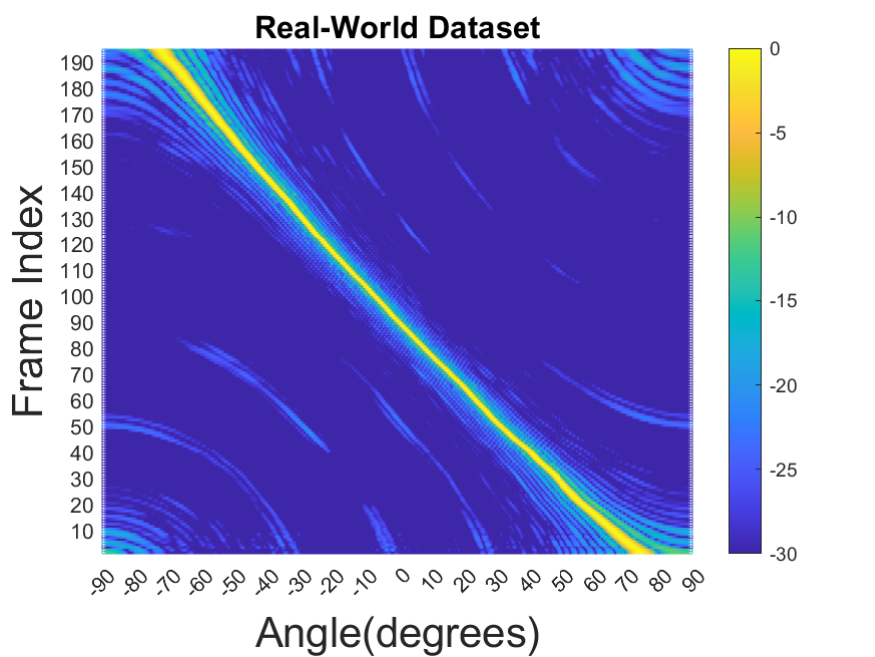}
\caption{ FFT spectrum of $195$ beam vectors, where the peak of the spectrum corresponds to the direction of targets. The color bar indicates the normalized magnitude in decibels (dB).} 
\label{rwdata}
\end{figure}

\subsubsection{Real-World Dataset}
Our field experiments utilized three multi-modal sensors: a Texas Instruments (TI) imaging radar, Teledyne FLIR Blackfly S stereo cameras, and Velodyne Ultra Puck VLP-32C LiDAR, as depicted in Figure \ref{plate}. The camera and LiDAR measurements served for experiment recording and provided the ground truth used to label the radar data. The TI cascaded imaging radar \cite{TI_Cascade} has the capability to synthesize an 86-element virtual uniform linear array with half-wavelength spacing along the horizontal direction. This is accomplished by employing MIMO radar technology with 9 transmit and 16 receive antennas.

The experiment was designed by parking a vehicle in an open parking lot, where a person holding a corner reflector walked slowly around the vehicle at a distance of approximately 15 meters. A total of $195$ beam vectors were selected from various angles to compose our real-world dataset. The angles of these beam vectors were estimated using FFT, as illustrated in Figure \ref{rwdata}. Notably, each beam vector exclusively contained a single target.

These individual beam vectors can serve as foundational components. To generate beam vectors with multiple targets, It is possible to superimpose beam vectors with different angles. Moreover, each beam vector comprised 86 elements. To align with our simulated data, a consecutive set of 64 random elements was chosen.

However, It is important to note that the SNR of this dataset remains constant. Due to constraints inherent to real-world conditions, this dataset is primarily employed here to showcase the illustrative performance of deep-MPDR, rather than for comprehensive statistical analysis.

\subsection{Training Approach}
The proposed deep-MPDR model underwent end-to-end training for $100$ epochs, employing a batch size of $1024$. The training process utilized the Adam optimizer with a learning rate set at $0.00001$, and the loss function selected was the mean squared error (MSE). The experiment itself was conducted within a Python 3.8 environment, utilizing PyTorch 1.10 and CUDA 11.1, all executed on four Nvidia RTX A6000 GPUs.

\begin{figure}[h]
\centering
\includegraphics[width= 3 in]{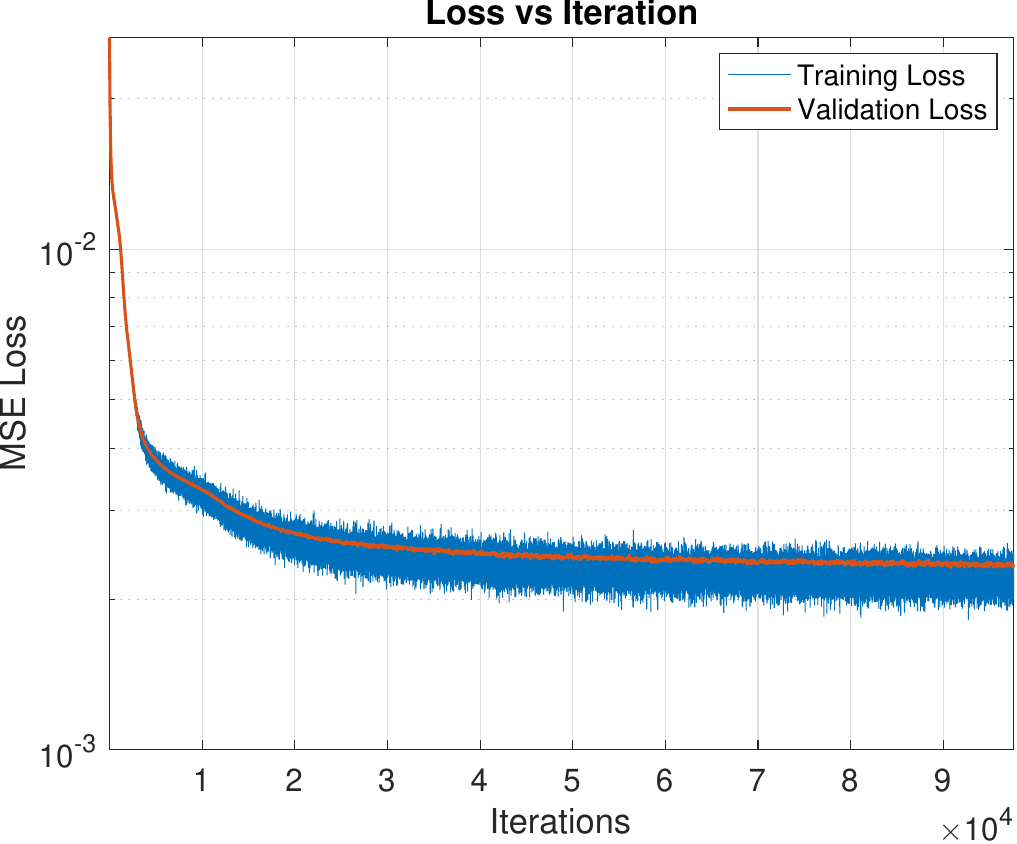}
\vspace{-0mm}
\caption{The training and validation loss of deep-MPDR. 
}
\label{loss}
\end{figure}

To counteract the risk of overfitting, a distinct validation set was created, mirroring the methodology of the training set but with distinct random reflection coefficients. Throughout the process, close attention was given to both the training and validation loss, depicted in Fig. \ref{loss}. The weight corresponding to the minimum validation loss was chosen as the foundation for all performance assessment tasks detailed in Section \ref{perf}.

\section{Performance Evaluation}\label{perf}
We assess the deep-MPDR model across four critical dimensions of DOA estimation: accuracy, separability, generalizability, and complexity. For a comprehensive comparison, we benchmark the deep-MPDR model against conventional DOA estimation techniques, including the IAA and Bartlett beamformers. 
We opt for the Bartlett beamformer in place of the MPDR beamformer due to an inherent dimensional mismatch between the covariance matrix after spatial smoothing and the signal.    

In our evaluation, we also include a convolutional neural network (CNN) \cite{papageorgiou2021deep} and a multi-layer perceptron (MLP) \cite{gall2020spectrum,gall2020learning}. These models are specifically designed for DOA estimation and framed as a multi-label classification task on a discrete grid. The CNN takes the covariance matrix as input and is structured with four 2D convolutional layers followed by four dense layers. The original CNN architecture is tailored for a 16-element Uniform Linear Array (ULA). When attempting to scale it up to a 64-element ULA while maintaining the same architecture, the parameter count increases significantly to $835,341,237$. In response, we introduce maxpooling layers between the convolutional layers to reduce the number of trainable parameters. The MLP includes an input layer that directly takes in the signal receive vector. This is succeeded by three hidden layers, each containing $256$, $512$, and $1024$ nodes, correspondingly. The activation function for the output layer has been switched to sigmoid. Both the CNN and MLP models are optimized using binary cross-entropy loss. 

\begin{figure}[h]
\centering
\includegraphics[width= 3.2 in]{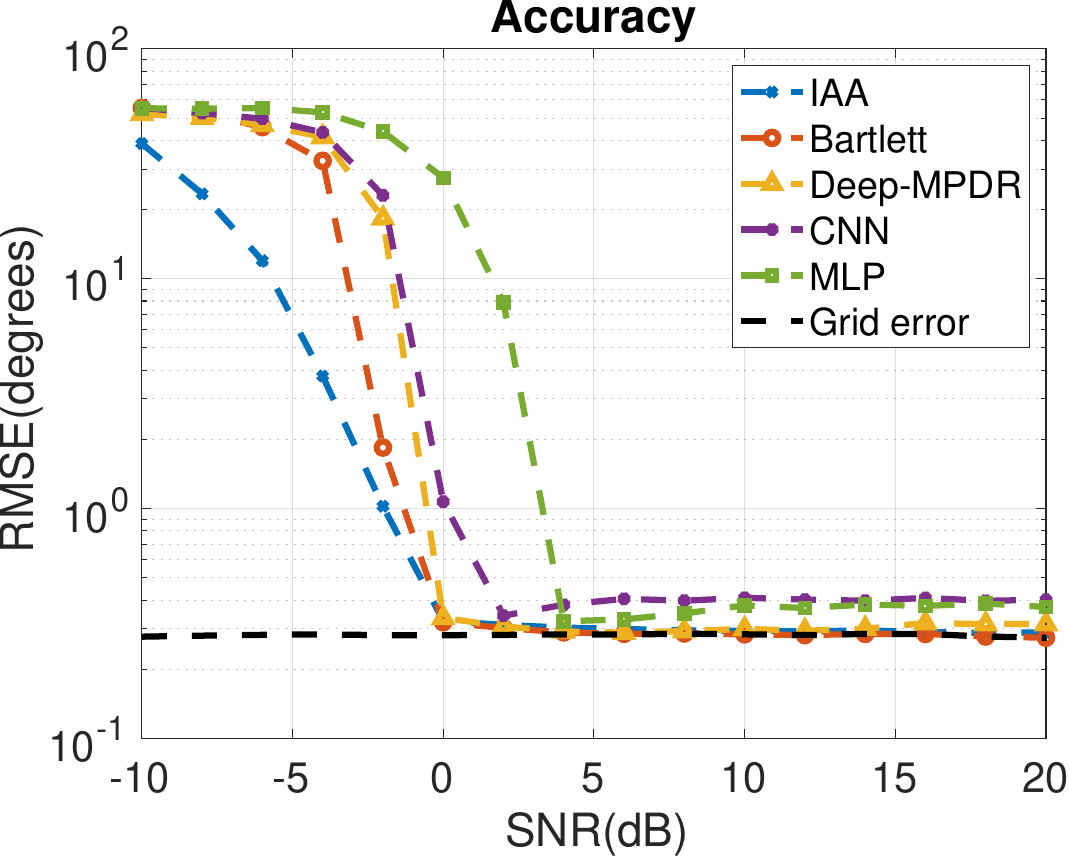}
\vspace{0mm}
\caption{The logarithmic scale RMSE versus SNR in the DOA estimation of a single, randomly generated off-grid target. The dark dashed line on the chart corresponds to the grid-induced error, which is computed as the RMSE between the source DOA and the nearest angle on the grid.}
\label{RMSE}
\end{figure}

Furthermore, to ensure consistency, all grid-based algorithms and deep learning models employ the same angle grid, discretizing the FOV from $-90^{\circ}$ to $90^{\circ}$ into $181$ equidistant points. To establish robustness, we conduct all experiments with $5,000$ Monte Carlo trials.

This comprehensive comparison aims to illuminate both the merits and limitations of the deep-MPDR model in contrast to these well-established methods. It also highlights the distinct advantages of the deep-MPDR model in single-snapshot DOA estimation.

\subsection{Accuracy}\label{acc}
We have opted for the root mean squared error (RMSE) as the primary performance metric to assess the accuracy of our DOA estimation methods. Our approach closely aligns with the standard grid-based DOA estimation methodology. This involves conducting a peak search to extract DOA estimates from the spectrum estimated by the deep-MPDR model. In each iteration of the Monte Carlo trial, a source off the grid is generated with a direction randomly selected from the range of $[-90^{\circ}, 90^{\circ}]$, along with an associated SNR.

 As illustrated in Figure \ref{RMSE}, the RMSE vs. SNR chart reveals that the deep-MPDR algorithm consistently outperforms both CNN and MLP models across all SNR scenarios. It delivers a comparable level of DOA estimation accuracy to that of IAA and Bartlett beamformer when the SNR is higher than $0$ dB. However, when the SNR falls below 0 dB, indicating that the noise power exceeds the signal power, the performance of all deep learning models experiences a significant drop.  

The observed phenomenon can be explained by the fact that all models were trained using a dataset with an SNR of 15dB and had no exposure to low SNR data. Incorporating low SNR data into the training dataset has the potential to improve model performance in noisy conditions. However, an excessive amount of noise in the dataset can introduce difficulties during model training and lead to decreased performance.

Unlike scenarios with multiple snapshots, where accurate covariance matrix estimation can still be achieved with multiple low SNR snapshots, in the case of a single snapshot, lower SNR primarily leads to a distorted beam pattern. Even with algorithms like IAA, its performance significantly deteriorates. When the RMSE exceeds a certain threshold, such as $5$ degrees, it becomes reasonable to consider the DOA estimation as inaccurate, and differing RMSE values do not necessarily reflect model performance. Moreover, in real-world scenarios, the SNR of single snapshot data typically exceeds 0 dB. For instance, in the case of automotive radar, the received signal benefits from a signal processing gain through the application of range-Doppler 2D FFTs\cite{Zheng_Asilomar_2022}.

\begin{figure}
\centering
\includegraphics[width= 3.2 in]{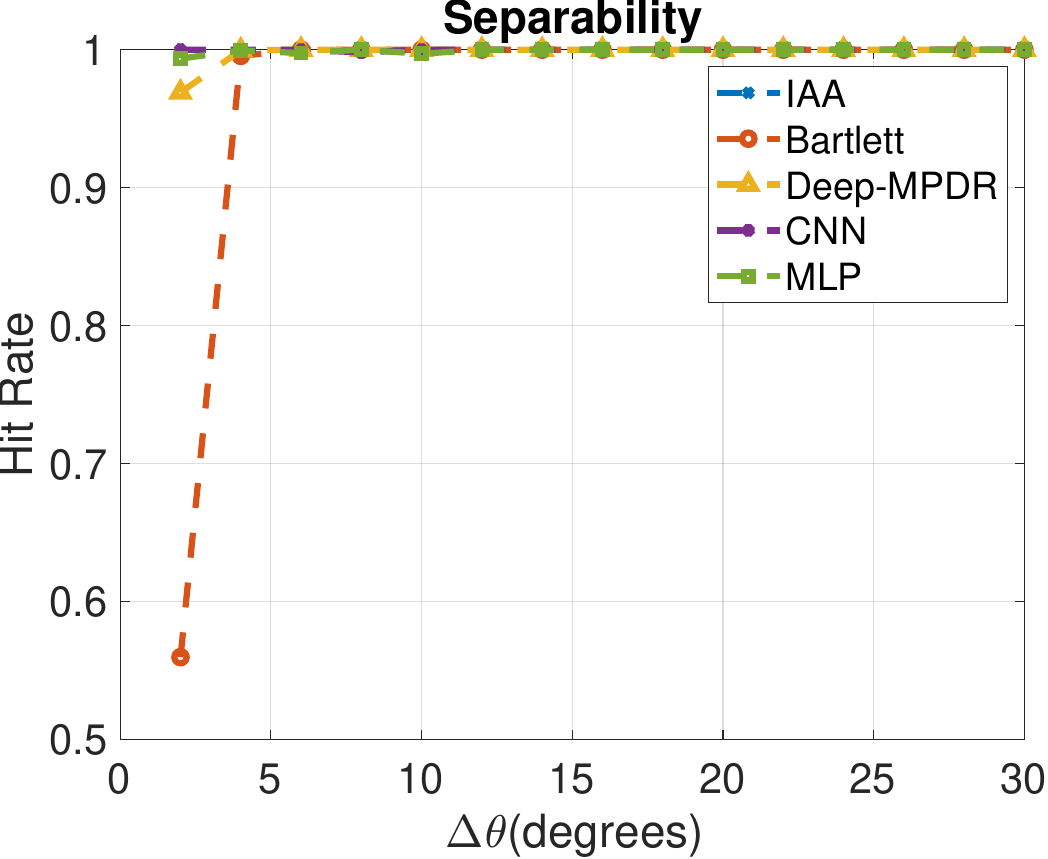}
\vspace{3mm}
\caption{Performance evaluation of IAA, Bartlett Beamformer, Deep-MPDR, CNN, and MLP:
hit rate vs $\Delta \theta$.}
\label{sep}
\end{figure}

\subsection{Separability}\label{separa}

To evaluate the effectiveness of DOA estimation in resolving closely situated targets, we conducted an experiment featuring two targets positioned at $-\Delta \theta / 2$ and $\Delta \theta / 2$, respectively, with $\Delta \theta$ representing the angular gap between them. A trial was considered successful when the deviation between estimated DOAs and actual values was within $\pm 1^{\circ}$. Hit rate calculation involved assessing the proportion of successful "hit" trials out of $5,000$ Monte Carlo trials conducted at an SNR of $15$ dB.

In Fig. \ref{sep}, all methods successfully resolve both targets when the separation is greater than $4^\circ$. Specifically, in the case of a $2^\circ$ separation between two targets, IAA achieves super-resolution and successfully detects both targets. The Bartlett beamformer's theoretical resolution is approximately $1.6^\circ$, determined by its 3-dB beamwidth for a 64-element ULA. However, the Bartlett beamformer's actual resolution is affected by the presence of noise, resulting in a $56\%$ hit rate for resolving both targets. Notably, all deep learning models demonstrate strong separability performance, with deep-MPDR slightly trailing the others when $\Delta \theta$ is $2^\circ$.

\subsection{Generalizability}
\subsubsection{K+1 Targets}
In order to extend the evaluation of the applicability of all deep learning models, an experiment was undertaken involving four off-grid targets positioned at $[-62.2^{\circ}, -21.9^{\circ},5.3^{\circ},45.1^{\circ}]$. Given that our training dataset is composed of scenarios only involving $K = 3$ targets, testing the models using data containing $K+1$ targets with diverse SNRs and randomly assigned reflection coefficients provides insight into the models' generalization capabilities.

\begin{figure}
\centering
\includegraphics[width= 3.2 in]{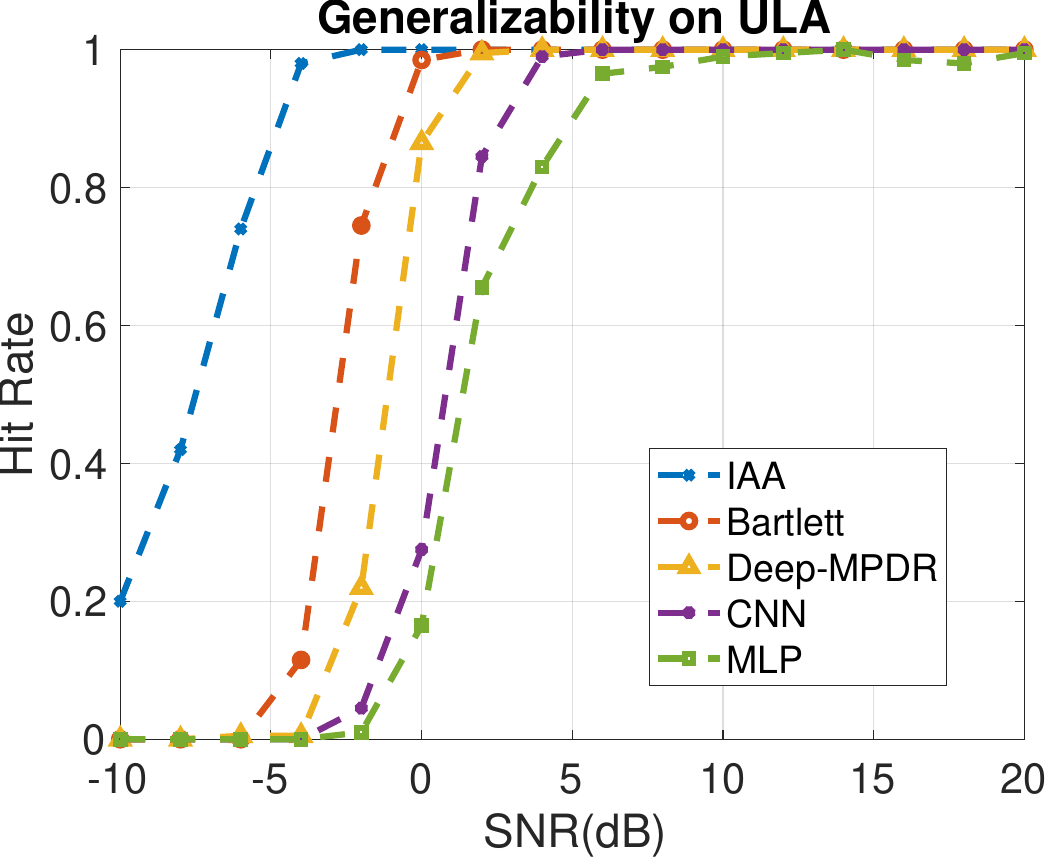}
\vspace{3mm}
\caption{K+1 Targets: Comparative Performance Evaluation of IAA, Bartlett Beamformer, Deep-MPDR, CNN, and MLP in Terms of Hit Rate versus SNR.}
\label{4p}
\end{figure}

\begin{figure}
\centering
\includegraphics[width= 3.2 in]{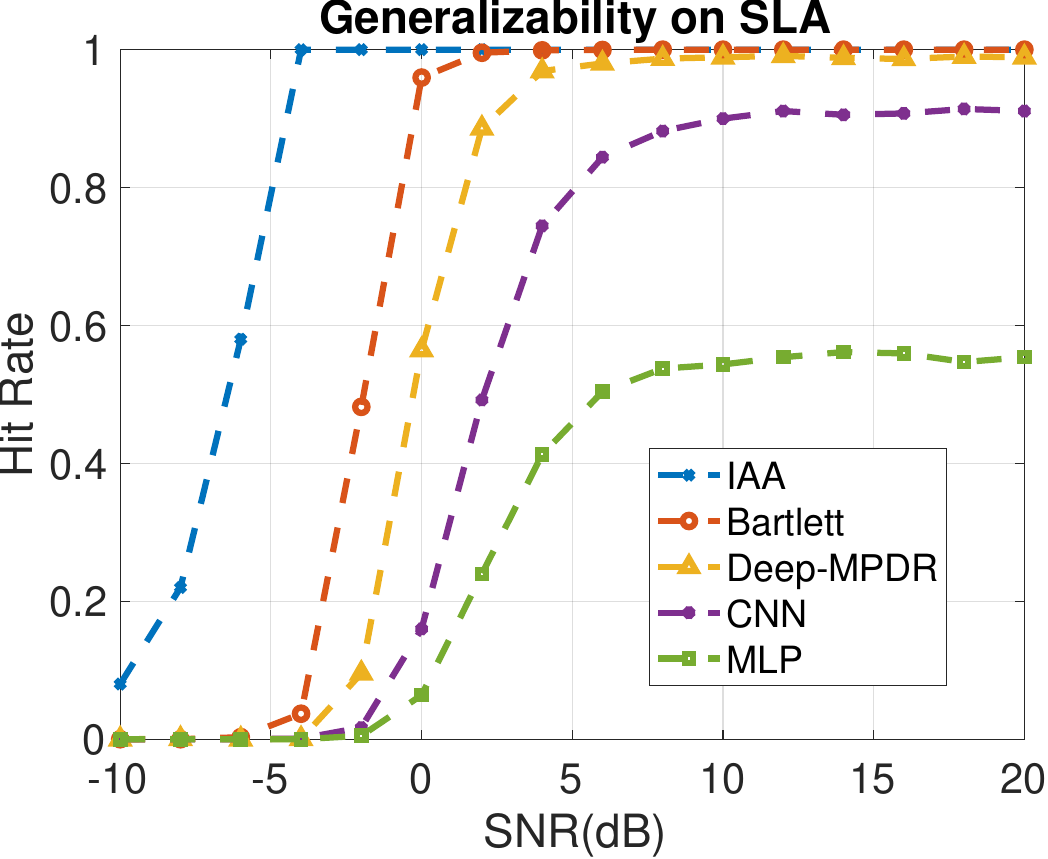}
\vspace{-0mm}
\caption{Sparse Arrays: Comparative Performance Evaluation of IAA, Bartlett Beamformer, Deep-MPDR, CNN, and MLP in Terms of Hit Rate versus SNR.}
\label{5p}
\end{figure}

\begin{figure*}
\centering
\subfloat[][$\boldsymbol{\Phi_1}$]{\includegraphics[width=0.33\linewidth]{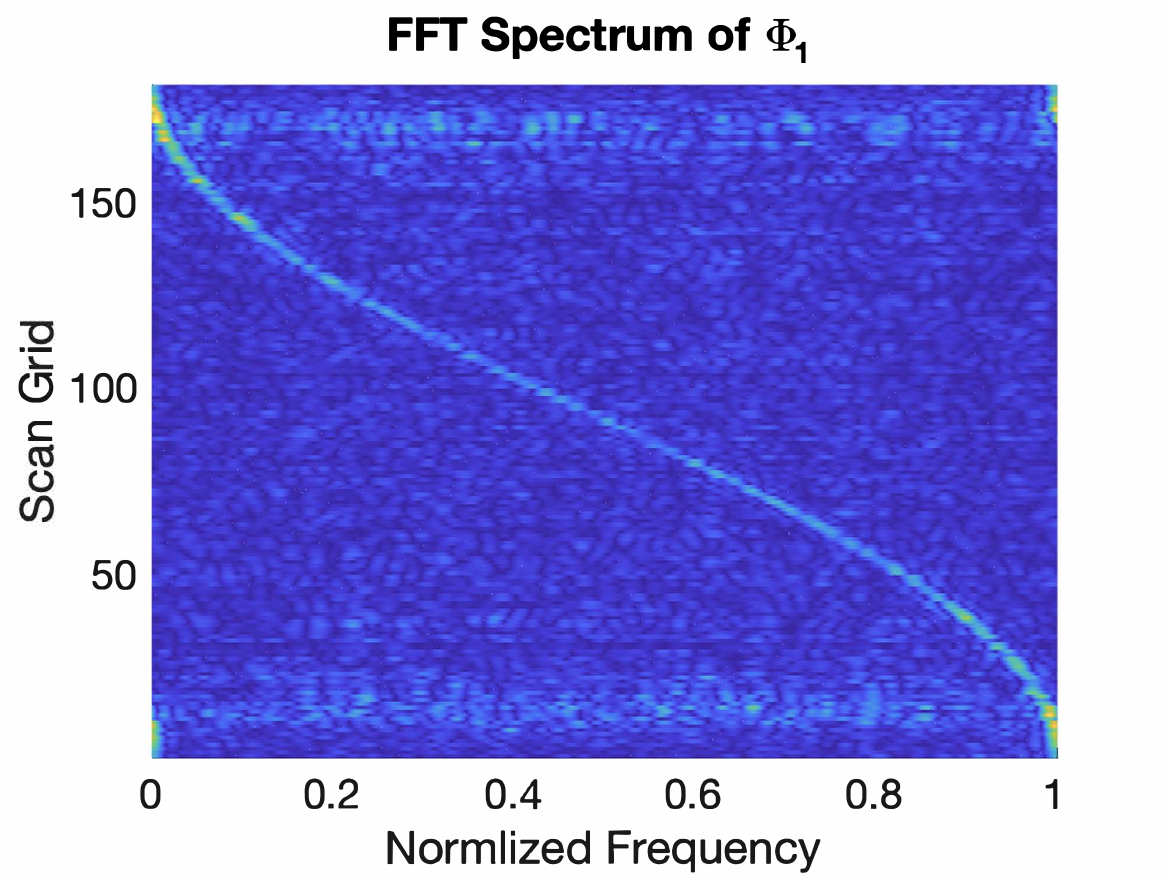}}
\subfloat[][$\boldsymbol{\Phi_2}$]{\includegraphics[width=0.33\linewidth]{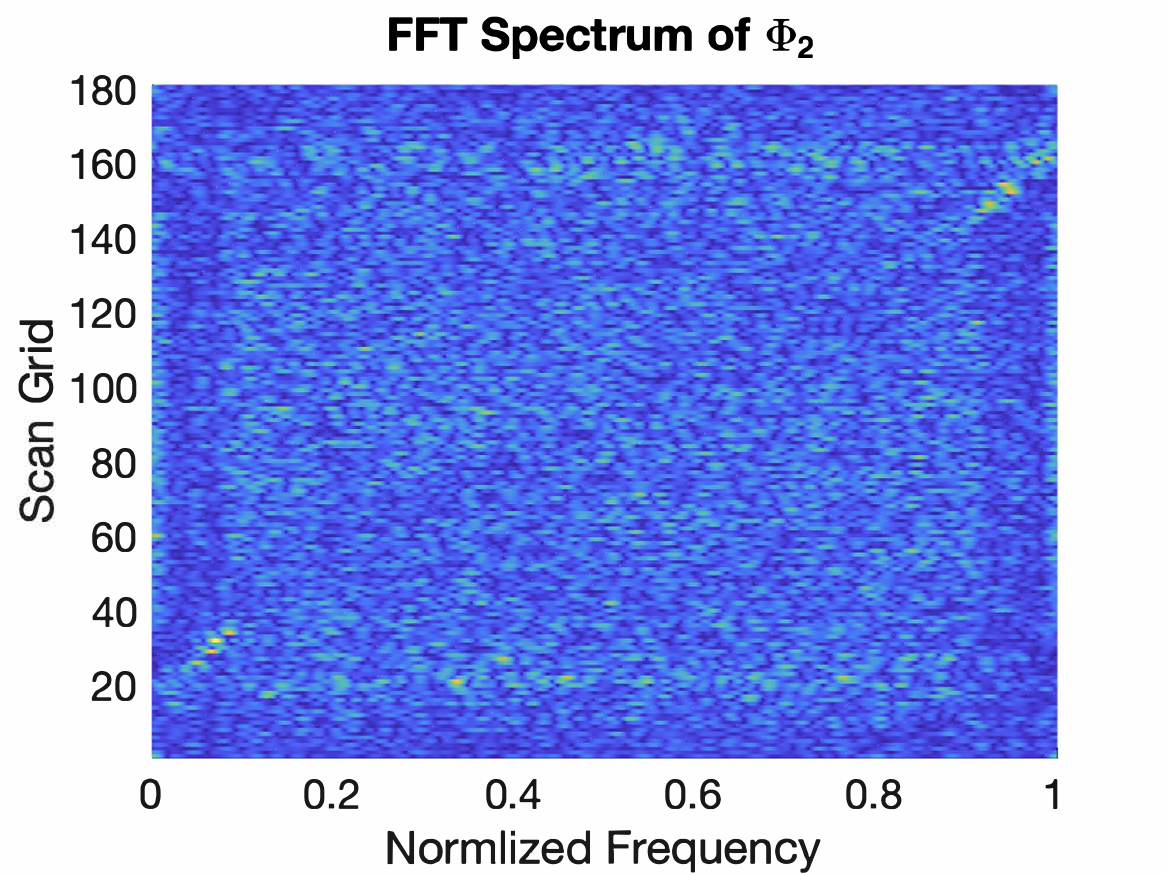}} 
\subfloat[][$\boldsymbol{\Phi_3}$]{\includegraphics[width=0.33\linewidth]{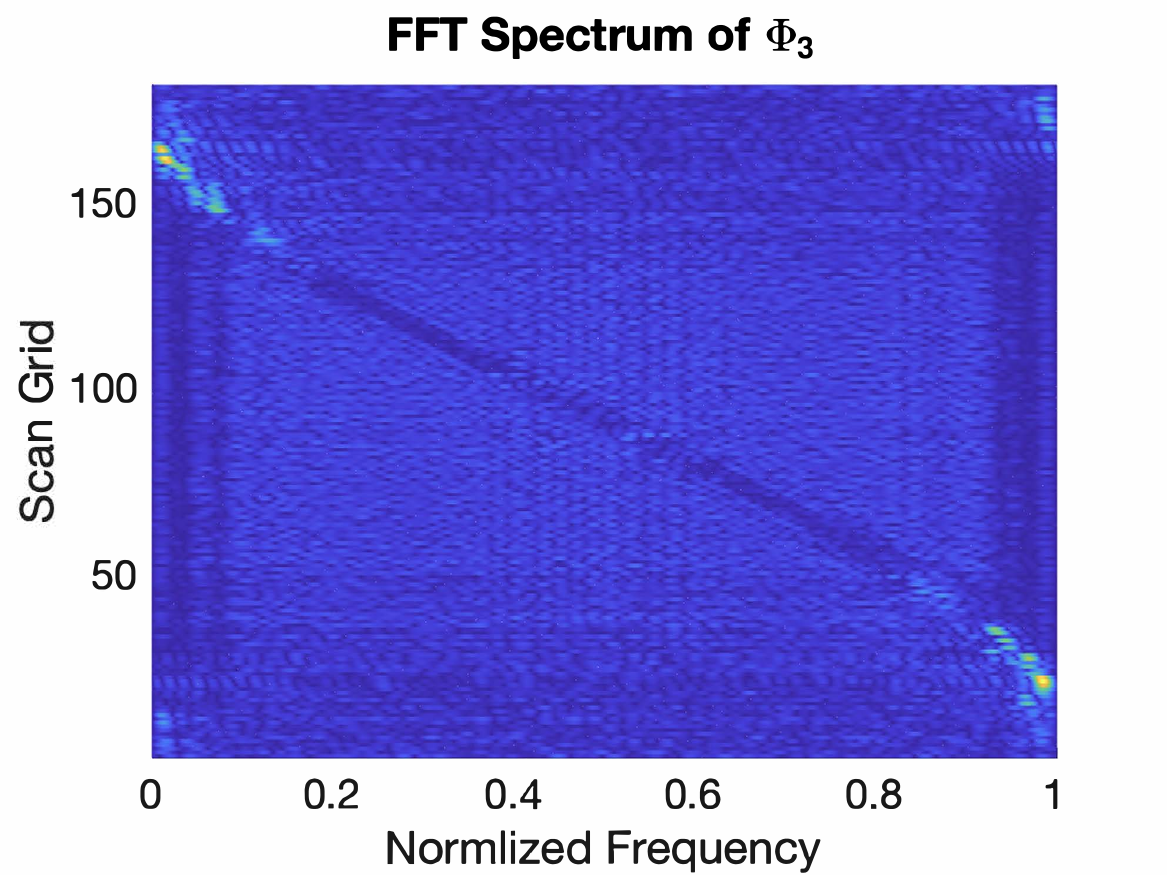}} \\
\subfloat[][$\boldsymbol{\Phi_4}$]{\includegraphics[width=0.33\linewidth]{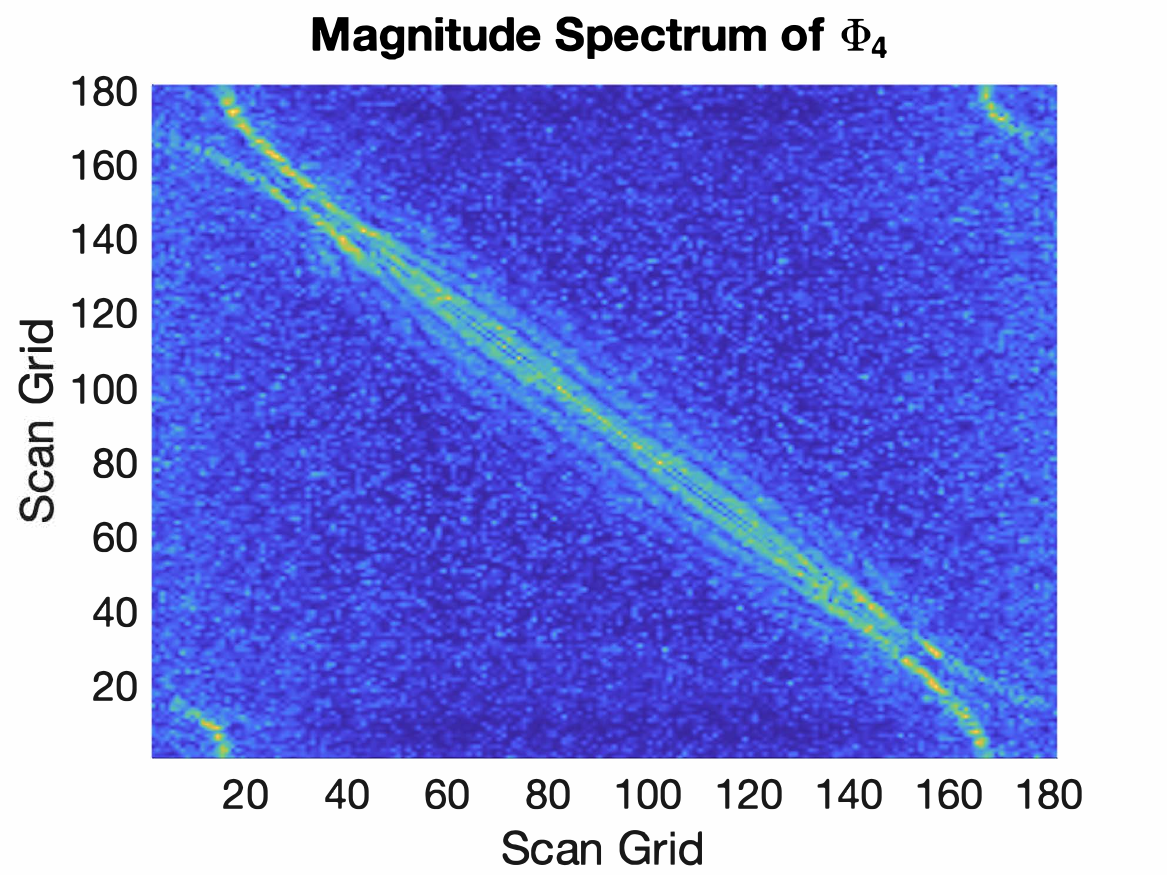}} 
\subfloat[][$\boldsymbol{\Phi_5}$]{\includegraphics[width=0.34\linewidth]{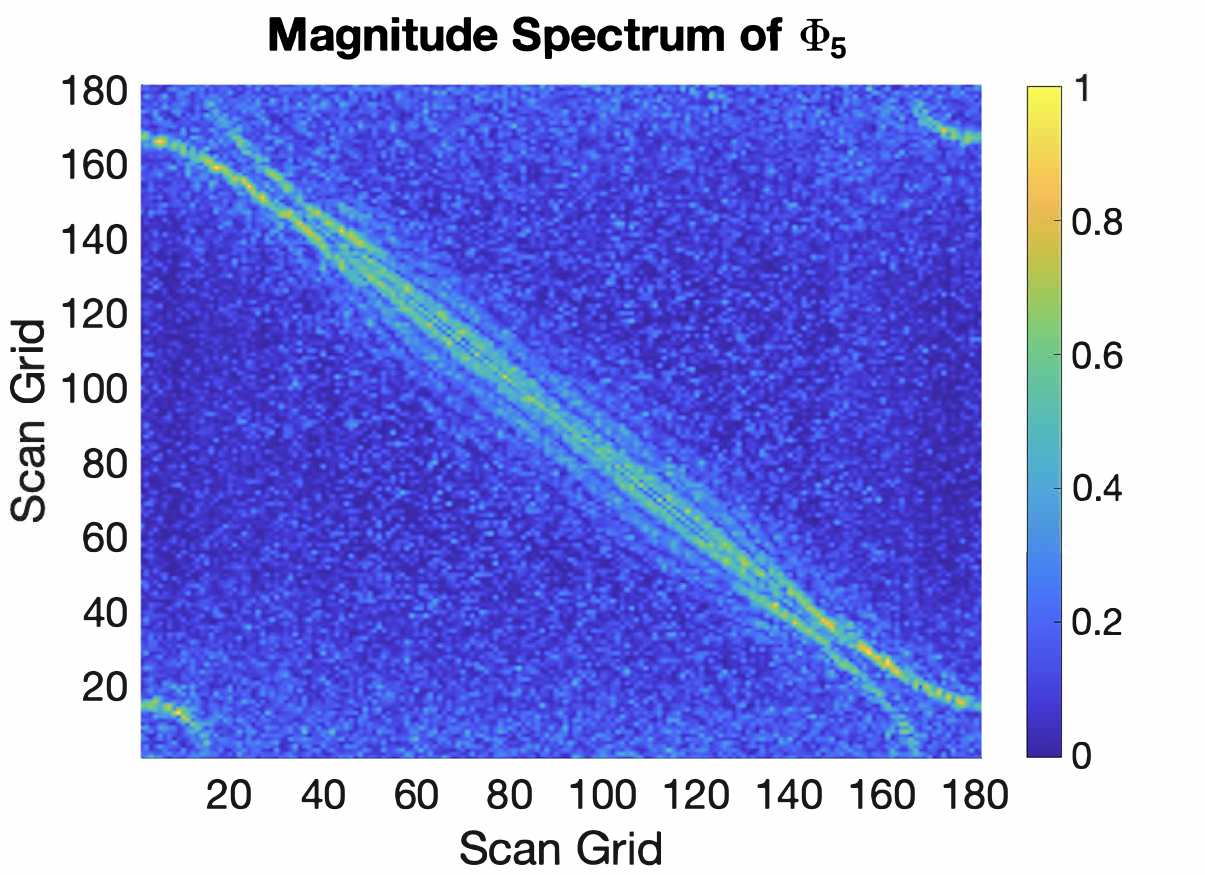}} 
\caption{FFT spectra and magnitude spectra of learnable parameters. All figures are plotted with normalized magnitude.} 
\label{parameters}
\end{figure*}

\subsubsection{Sparse Arrays} \label{sparse}
In the practical implementation of deep learning models for DOA estimation tasks, the capability to handle sparse array structures holds significant importance. Sparse arrays can serve the purpose of deliberate cost reduction and mitigation of mutual coupling effects. Alternatively, they might emerge due to antenna element failures within a ULA. Retraining models each time for different array structures is impractical. Therefore, as part of our generalizability experiment, we assess model performance on randomly created sparse arrays. In each trial, $50$ elements are randomly selected out of the 64-element ULA, while maintaining a consistent aperture setup, where the first element is positioned at location $1$ and the last element at location $64$. The signal at unselected elements is set to zero to align with the input layer of deep learning models. Apart from this adjustment, all other experimental conditions remain unchanged.

The results of the generalizability experiment are depicted in Fig. \ref{4p} and \ref{5p}, where the performance of IAA, Bartlett beamformer, and three deep learning models was compared. Deep-MPDR achieves a similar high hit rate to model-based algorithms when SNR is greater than 0 dB. However, under low SNR scenarios, all three deep learning models perform worse than model-based algorithms, which is consistent with the results shown in Section \ref{acc}.

On the other hand, Deep-MPDR demonstrates a better hit rate in both the K+1 targets test and sparse arrays test across all SNR scenarios due to its model-based nature, providing better generalizability than conventional deep learning methods. CNN outperforms MLP in both tests because of its larger number of parameters, nearly three times that of MLP. Overall, Deep-MPDR exhibits good generalizability compared to other deep learning models, highlighting its potential for broader applications and real-world usage

\subsection{Complexity}
We conducted a comprehensive assessment of the computational complexity of the deep-MPDR model by analyzing its inference time and the number of trainable parameters. To ensure fairness, all DOA methods were executed in Python 3.8 using PyTorch 1.10 and CUDA 11.1 on a single Nvidia RTX A6000 GPU, and the inference time was determined by averaging over $5,000$ trials. For the deep learning-based methods, a batch size of 1 was employed. The results, as presented in Table \ref{table_1}, highlight that the Bartlett beamformer exhibits the fastest inference time, primarily because it requires only a single matrix multiplication operation. In contrast, IAA has the longest inference time due to its iterative matrix inversion process, rendering it unsuitable for real-time implementation. Among the deep learning models, the CNN method has a larger number of trainable parameters because it incorporates 2D convolutional layers specifically designed for processing the 2D covariance matrix input. Meanwhile, MLP proves slightly faster than deep-MPDR, mainly because it consists of only three fully connected layers. Remarkably, deep-MPDR employs the fewest parameters among the three deep learning models, highlighting its parameter efficiency.
\begin{table}[h]
\centering
\resizebox{\linewidth}{!}{%
\begin{tabular}{|l|c|c|} \hline 
\textbf{Methods} & \textbf{Inference Time (ms)} & \textbf{\# Trainable Parameters} \\ \hline

Bartlett  & $0.014$ & --\\

IAA  & $686.20$  & -- \\

CNN  & $1.26$ & $2,109,877$  \\

MLP  & $0.63$ & $875,445$  \\

Deep-MPDR  & $0.89$ & $223,718$ \\\hline
\end{tabular}}
\vspace{3 mm}
\caption{Inference Time Comparison of DOA Methods}
\label{table_1}
\end{table}

\begin{figure*}
\centering
\subfloat[][MPDR]{\includegraphics[width=0.33\linewidth]{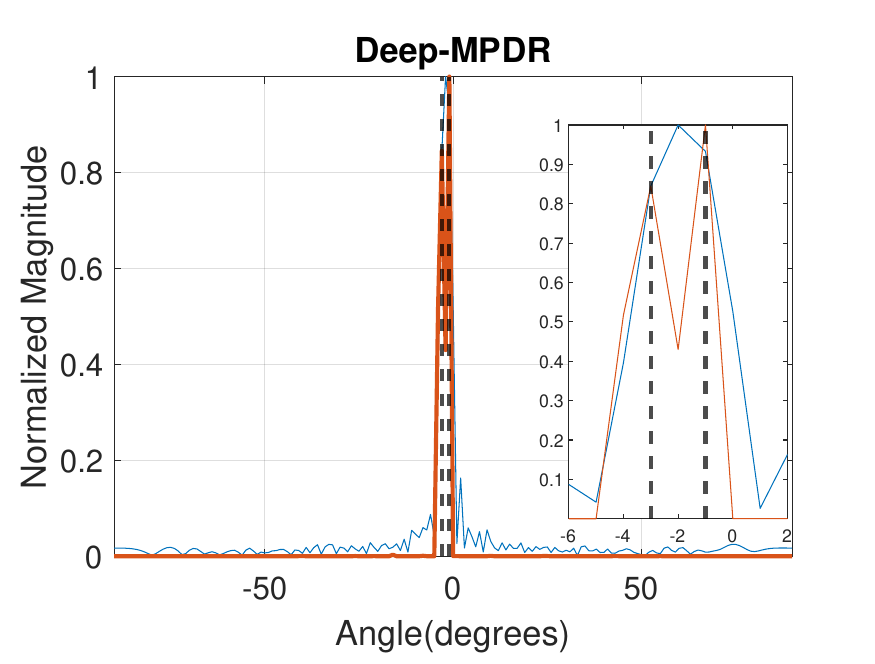}}
\subfloat[][CNN]{\includegraphics[width=0.33\linewidth]{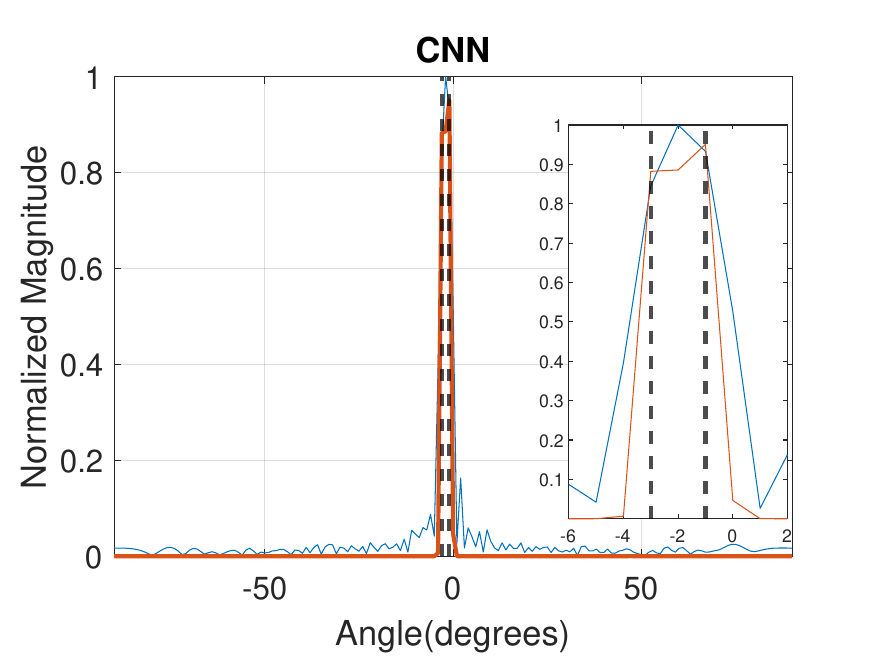}} 
\subfloat[][MLP]{\includegraphics[width=0.33\linewidth]{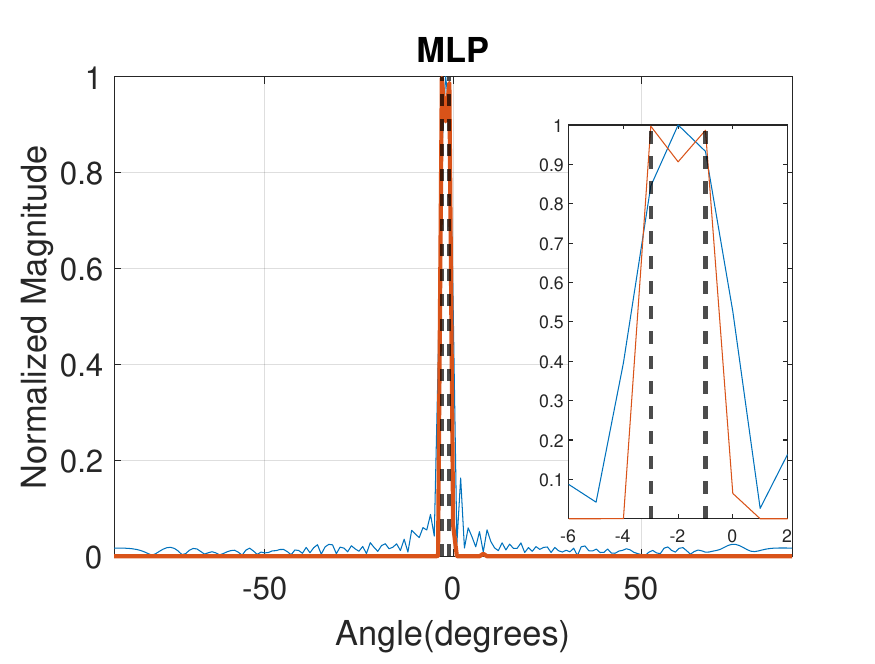}}\\
\subfloat[][MPDR]{\includegraphics[width=0.33\linewidth]{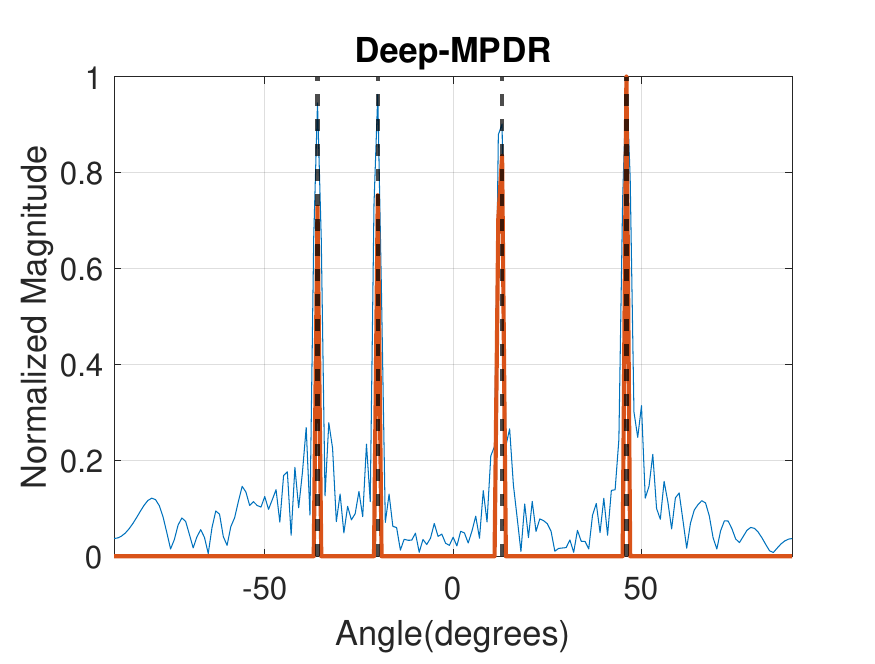}}
\subfloat[][CNN]{\includegraphics[width=0.33\linewidth]{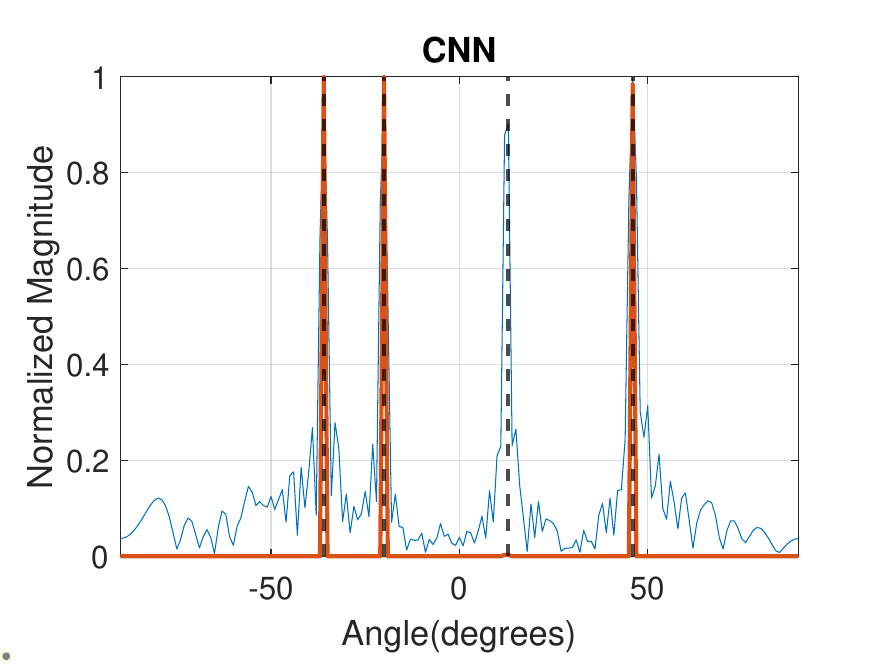}} 
\subfloat[][MLP]{\includegraphics[width=0.33\linewidth]{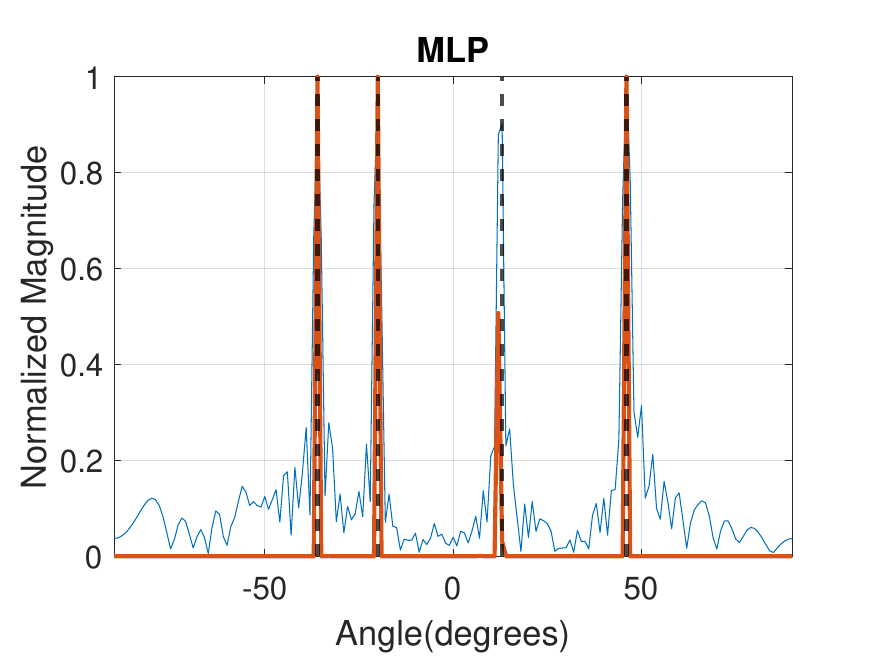}}\\
\subfloat[][MPDR]{\includegraphics[width=0.33\linewidth]{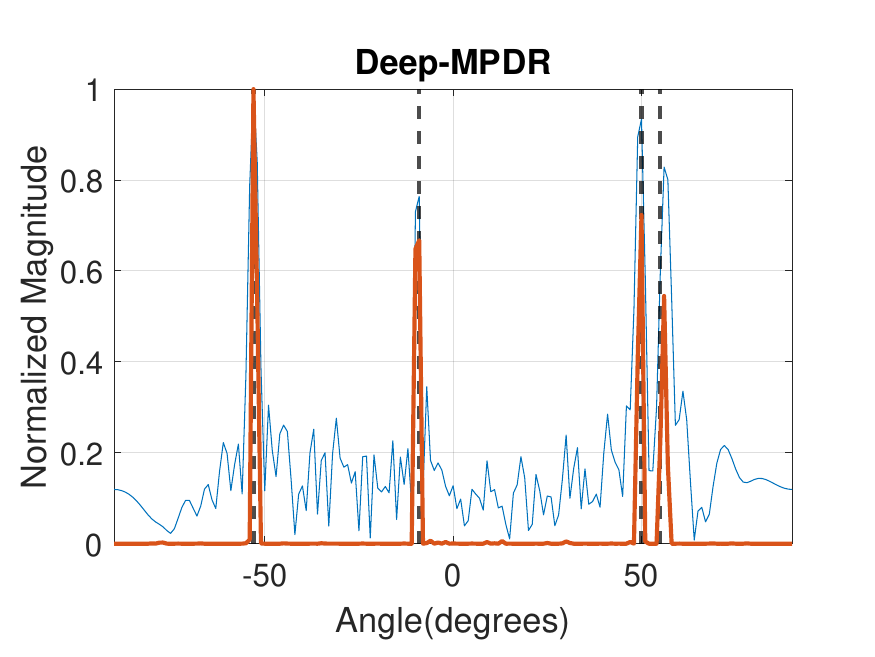}}
\subfloat[][CNN]{\includegraphics[width=0.33\linewidth]{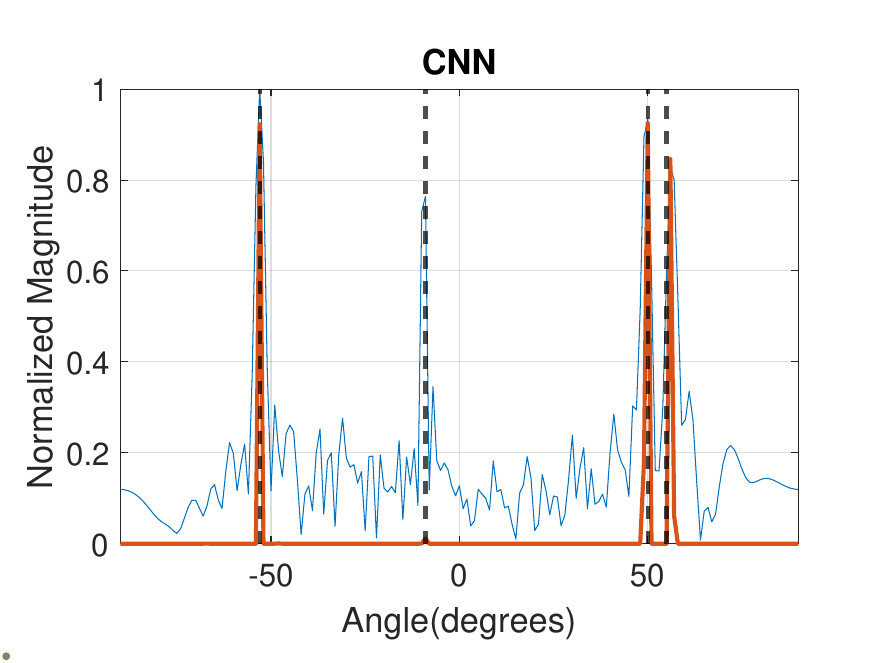}} 
\subfloat[][MLP]{\includegraphics[width=0.33\linewidth]{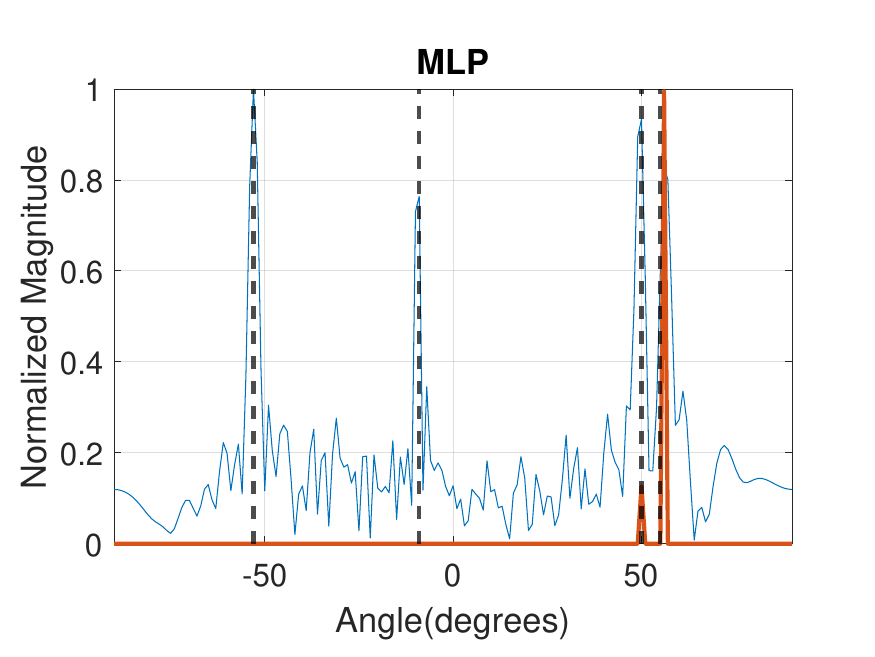}}
\caption{Detection results of deep-learning models on real-world data for three scenarios. Blue lines represent spectra generated by the Bartlett beamformer, red lines represent spectra generated by corresponding deep-learning models, and the black dashed line represents the ground truth DOA generated by IAA on an 86-element signal.} 
\label{rea_data}
\end{figure*}

\vspace{-3mm}

\subsection{Interpretability}
Beyond performance, interpretability is a critical aspect of deep-learning models. Unlike CNN and MLP models, which can be challenging to understand solely by examining the network parameters, we can establish the interpretability of the deep-MPDR model through a comprehensive analysis of its parameters.

From section \ref{arch}, we know that $\boldsymbol{\Phi}_1$, $\boldsymbol{\Phi}_2$, and $\boldsymbol{\Phi}_3$ are linked to the dictionary matrix $\textbf{A}$. Through FFT operations on each scan grid, FFT spectra are generated. Fig. \ref{parameters}a displays the FFT spectrum of $\boldsymbol{\Phi}_1$, revealing similarities to a dictionary matrix $\boldsymbol{A}$. Furthermore, Fig. \ref{parameters}b and Fig. \ref{parameters}c show the FFT spectra of $\boldsymbol{\Phi}_2$ and $\boldsymbol{\Phi}_3$, respectively, both indicating structural characteristics in the frequency domain. Furthermore, Fig. \ref{parameters}d and Fig. \ref{parameters}e display the magnitude spectra, obtained by taking the absolute values of the complex components, of $\boldsymbol{\Phi}_4$ and $\boldsymbol{\Phi}_5$, respectively, demonstrating that both parameters predominantly resemble diagonal-dominant matrices.

\subsection{Real-World Data Examples}
In this section, we show some detection examples of deep learning models on real-world data. The ground truth of the real-world data is generated using IAA on 86-element ULA, and a consecutive 64-element ULA is extracted from the original 86-element ULA to serve as the input for the deep learning models. 

Figures \ref{rea_data}a, \ref{rea_data}b, and \ref{rea_data}c depict scenarios with two closely placed targets at directions of $-3^\circ$ and $-1^\circ$. Upon closer examination of these zoomed-in figures, it becomes evident that the Bartlett beamformer is incapable of resolving these two targets. On the other hand, all deep-learning models successfully resolve the two targets, with our deep-MPDR model producing sharper detection peaks compared to other deep-learning models.

Figures \ref{rea_data}d, \ref{rea_data}e, and \ref{rea_data}f depict scenarios with $K+1$ targets placed at directions of $-36^\circ$, $-20^\circ$, $13^\circ$, and $46^\circ$, respectively. Leveraging the advantages of the large aperture offered by the 64-element ULA, the Bartlett beamformer successfully detects all four targets, albeit with notable high sidelobe levels. Among the deep learning models, our deep-MPDR demonstrates its superior generalizability by effectively detecting all four targets with prominent peaks. However, the CNN model fails to detect one target at $13^\circ$, and while the MLP model successfully identifies all four targets, the peak at $13^\circ$ is relatively subdued.

Figures \ref{rea_data}g, \ref{rea_data}h, and \ref{rea_data}i illustrate scenarios with $K+1$ targets positioned at directions of $-53^\circ$, $-9^\circ$, $50^\circ$, and $55^\circ$, respectively, utilizing a sparse array. The sparse array is generated using the same procedure as described in Section \ref{sparse}. While the Bartlett beamformer can detect all four targets, the sidelobe levels are elevated due to the sparse configuration. Our deep-MPDR model demonstrates its remarkable generalizability by successfully detecting all four targets, although there is a slight drop in the peak at $55^\circ$. In contrast, both CNN and MLP models struggle to detect all the targets, with CNN failing to detect the target at $-9^\circ$, and MLP only detecting one target at $55^\circ$.

\section{Conclusions}
By harnessing the strengths of both the MPDR beamformer and recent deep learning-based DOA estimation techniques, our deep-MPDR network transforms the MPDR beamformer into a deep-learning framework. Through the integration of domain knowledge, our deep-MPDR model utilizes fewer parameters while providing an interpretable and efficient solution for high-resolution angle finding in single snapshot scenarios. It achieves fast inference times by circumventing the high computational cost associated with large matrix inversions and exhibits superior generalization capabilities compared to purely data-driven deep learning approaches. We demonstrate the superior performance of our deep-MPDR network through extensive numerical experiments on both simulated and real-world datasets. This innovative approach demonstrates strong performance in controlled settings and holds promise for real-time DOA estimation applications in automotive radar scenarios.

\bibliographystyle{IEEEtran}
\bibliography{ref}

\begin{IEEEbiography}[{\includegraphics[width=1in,height=1.25in,clip,keepaspectratio]{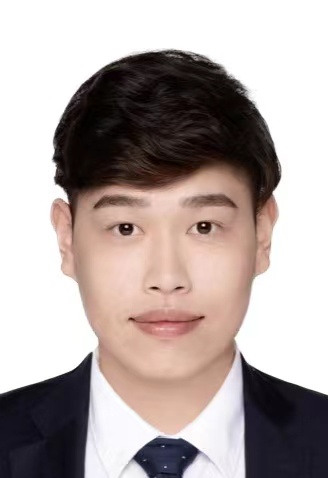}}]%
{Ruxin Zheng} (Student Member, IEEE) 
received his B.S. degree in Electrical Engineering from the University of Pittsburgh, Pittsburgh, PA, USA, in 2018, and his M.S. degree in Electrical Engineering from the University of Michigan, Ann Arbor, MI, USA, in 2020. Currently, He is working towards his Ph.D. degree at the Department of Electrical and Computer Engineering at The University of Alabama, Tuscaloosa, AL, USA. His research interests are automotive radar, radar signal processing, MIMO radar with sparse sensing, and machine learning. 
\end{IEEEbiography}

\begin{IEEEbiography}[{\includegraphics[width=1in,height=1.25in,clip,keepaspectratio]{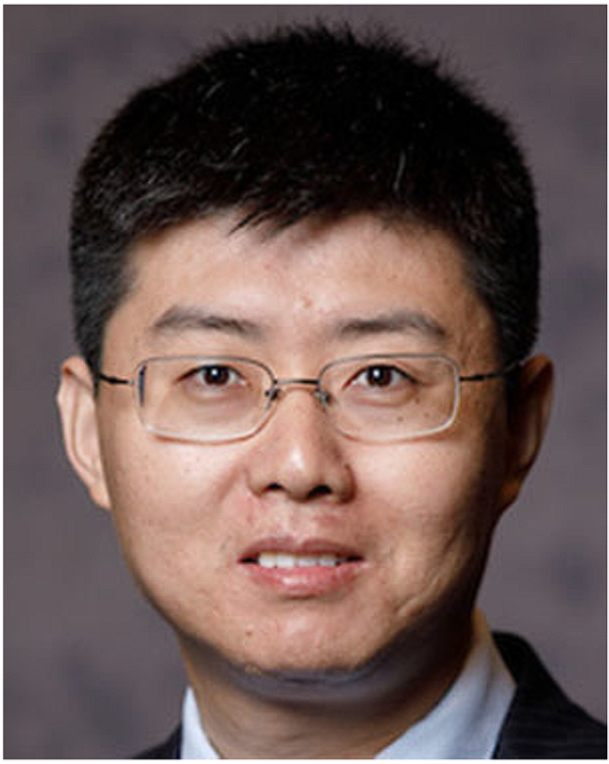}}]%
{Shunqiao Sun}
(Senior Member, IEEE) received the Ph.D. degree in Electrical and Computer Engineering from Rutgers, The State University of New Jersey, New Brunswick, NJ, USA, in January 2016. 

Dr. Sun is currently a tenure-track Assistant Professor at the Department of Electrical and Computer Engineering, The University of Alabama, Tuscaloosa, AL, USA. His research interests lie at the interface of statistical and sparse signal processing with mathematical optimizations, automotive radar, MIMO radar, machine learning, and smart sensing for autonomous vehicles. From 2016-2019, he was with the radar core team of Aptiv, Technical Center Malibu, California, where he has worked on advanced radar signal processing and machine learning algorithms for self-driving vehicles and lead the development of DOA estimation techniques for next-generation short-range radar sensor which has been used in over 120-million automotive radar units.  

Dr. Sun has been awarded 2016 IEEE Aerospace and Electronic Systems Society Robert T. Hill Best Dissertation Award for his thesis ``MIMO radar with Sparse Sensing''. He authored a paper that won the Best Student Paper Award at 2020 IEEE Sensor Array and Multichannel Signal Processing Workshop (SAM). He is an elected member of IEEE Sensor Array and Multichannel (SAM) technical committee (2024-2026), Vice Chair of IEEE Signal Processing Society Autonomous Systems Initiative (ASI) Steering Committee (2023-2024). He is an Associate Editor of IEEE Signal Processing Letters and IEEE Open Journal of Signal Processing. He is a Senior Member of IEEE.
\end{IEEEbiography}

\begin{IEEEbiography}[{\includegraphics[width=1in,height=1.25in,clip,keepaspectratio]{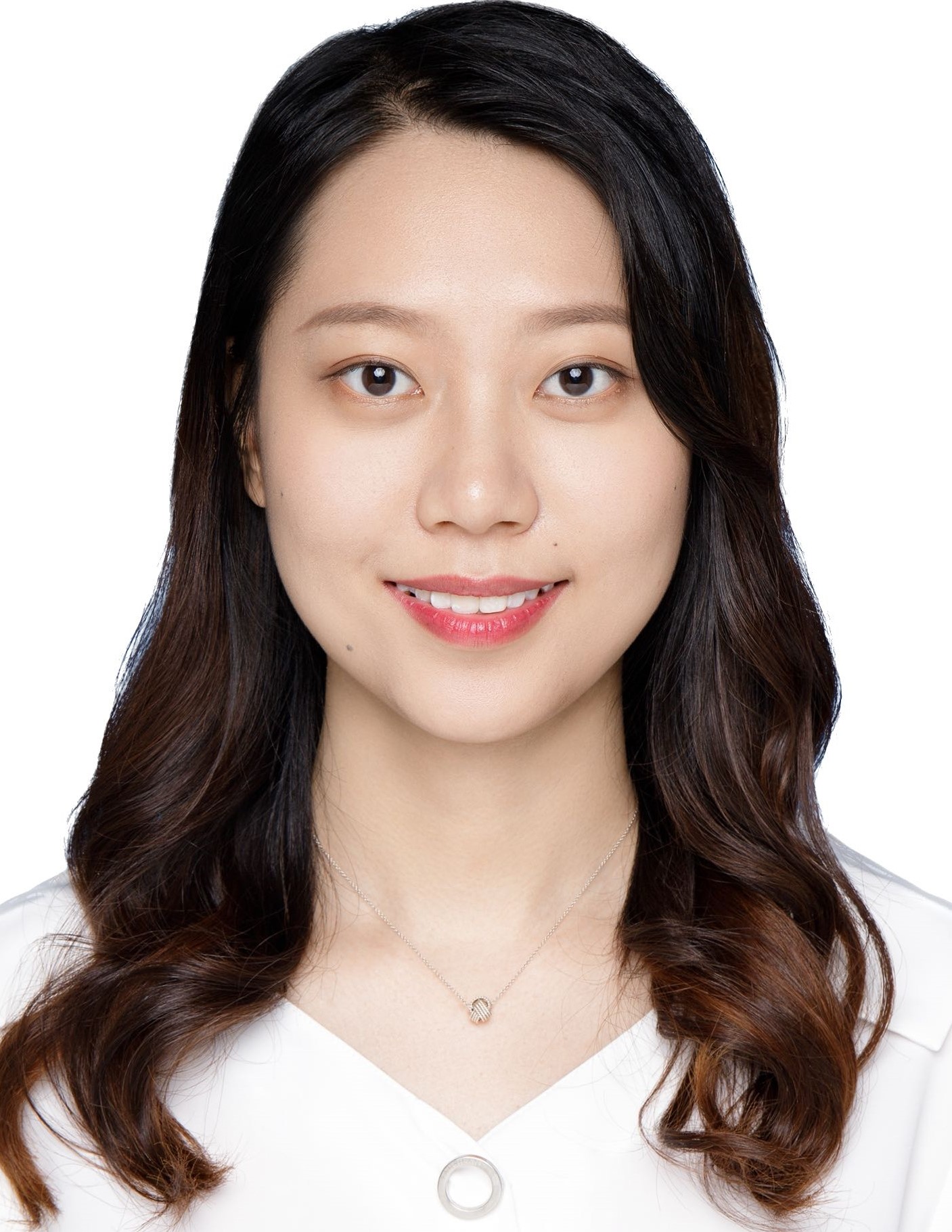}}]%
{Hongshan Liu} (Student Member, IEEE)
is a doctoral student in Electrical Engineering at The University of Alabama. Before that, she received her M.S. degree in Electrical Engineering from the University of Michigan-Ann Arbor and her B.S. degree in Physics from Zhejiang University. Her research focuses on deep learning-based image processing.
\end{IEEEbiography}

\begin{IEEEbiography}[{\includegraphics[width=1in,height=1.25in,clip,keepaspectratio]{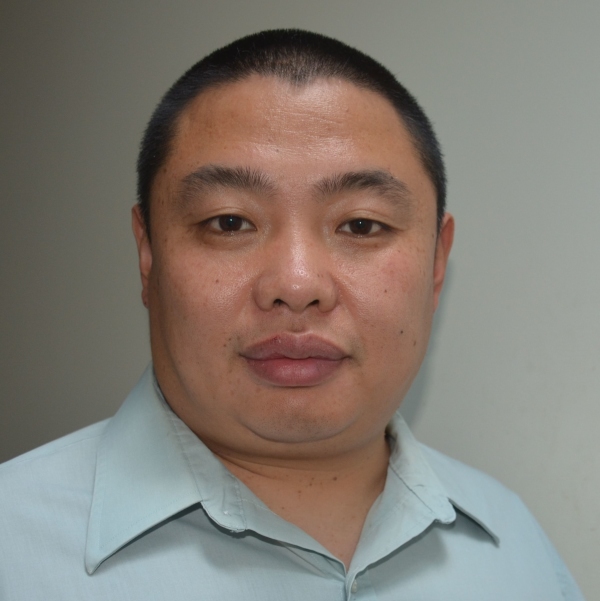}}]%
{Honglei Chen} (Member, IEEE) received his B.S. degree from Beijing Institute of Technology and his M.S. and Ph.D. degrees from University of Massachusetts Dartmouth,  all in Electrical Engineering. 

Dr. Chen is a principal engineer at MathWorks where he leads the development of phased-array system simulation tools and algorithms for radar, 5G, sonar, and ultrasound applications.  
\end{IEEEbiography}

\begin{IEEEbiography}[{\includegraphics[width=1in,height=1.25in,clip,keepaspectratio]{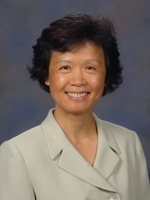}}]%
{Jian Li} (Fellow, IEEE)
received the M.Sc. and Ph.D. degrees in electrical engineering from The Ohio
State University, Columbus, in 1987 and 1991, respectively. 

She is currently a Professor
in the Department of Electrical and Computer Engineering, University of Florida,
Gainesville. Her current research interests include spectral estimation, statistical and
array signal processing, and their applications to radar, sonar, and biomedical engineering.
Dr. Li's publications include Robust Adaptive Beamforming (2005, Wiley), Spectral
Analysis: the Missing Data Case (2005, Morgan \& Claypool), MIMO Radar Signal
Processing (2009, Wiley), and Waveform Design for Active Sensing Systems -- A
Computational Approach (2011, Cambridge University Press).

Dr. Li is a Fellow of IEEE and a Fellow of IET. She is a Fellow of the European
Academy of Sciences (Brussels). She received the 1994 National Science Foundation
Young Investigator Award and the 1996 Office of Naval Research Young Investigator
Award. She was an Executive Committee Member of the 2002 International Conference
on Acoustics, Speech, and Signal Processing, Orlando, Florida, May 2002. She was an Associate Editor of the IEEE Transactions on Signal Processing from 1999 to 2005, an
Associate Editor of the IEEE Signal Processing Magazine from 2003 to 2005, and a
member of the Editorial Board of Signal Processing, a publication of the European
Association for Signal Processing (EURASIP), from 2005 to 2007. She was a member of
the Editorial Board of the IEEE Signal Processing Magazine from 2010 to 2012. She is a
co-author of the paper that has received the M. Barry Carlton Award for the best paper
published in IEEE Transactions on Aerospace and Electronic Systems in 2005. She is
also a co-author of a paper published in IEEE Transactions on Signal processing that has
received the Best Paper Award in 2013 from the IEEE Signal Processing Society.
\end{IEEEbiography}

\end{document}